\documentclass[twocolumn,tighten,times]{aastex63}
\usepackage[T1]{fontenc}

\let\tablenum\relax 
\usepackage{siunitx}
\usepackage{appendix}

\input{newcommands.sty} 

\received{July 1,2020}
\revised{October 2, 2020; November, 5th 2020}
\accepted{November 12, 2020}

\submitjournal{ApJ}

\shorttitle{Testing Jeans, Toomre and Bonnor-Ebert concepts for Planetesimals.}
\shortauthors{Klahr and Schreiber}

\newcommand{\konsti}[1]{{#1}}

\newcommand{\correctedn}[1]{#1}
\newcommand{\corrected}[1]{{#1}}

\begin{document}
\title{Testing the Jeans, Toomre and Bonnor-Ebert concepts for planetesimal formation:\\
3D streaming instability simulations of diffusion regulated formation of planetesimals}

\correspondingauthor{Hubert Klahr}
\email{klahr@mpia.de}

\author[0000-0002-8227-5467]{Hubert Klahr}
\author{Andreas Schreiber}
\affil{Max Planck Institut f\"ur Astronomie, K\"onigstuhl 17, 69117, Heidelberg, Germany}



\begin{abstract}
We perform streaming instability simulations at Hill density and beyond, to demonstrate that Planetesimal formation is not completed when pebble accumulations exceed the local Hill density. We find that Hill density is not a sufficient criterion for further gravitational collapse of a pebble cloud into a planetesimal, but that additionally the accumulated mass has to be large enough to overcome turbulent diffusion.
A Toomre analysis of the system indicates that linear self-gravity modes play no role on the scale of our numerical simulation. We nevertheless find that self-gravity, by vertically contracting the pebble layer, increases the strength of turbulence, which is either an indication of Kelvin Helmholtz Instability or a boost of the streaming-instability. We furthermore determine the Bonnor-Ebert central density to which a pebble cloud of given mass has to be compressed before it would be able to continue contraction against internal diffusion. As the equivalent "solid body" size of the pebble cloud scales with the central density to the power of -1/6, it is much easier to have a pebble cloud of 100 km equivalent size to collapse than one of 10 km for the same level of turbulent diffusion. This can explain the lack of small bodies in the solar system and predicts to have small objects formed by gravitational collapse at large pebble to gas ratios, in the outskirts of the solar nebula and at late times of generally reduced gas mass.
\end{abstract}


\section{Introduction} \label{sec:intro}
Two ways are currently known to form planetesimals in the solar nebula or more generally in a protoplanetary disk.
Either pebbles grow via sticking collisions to larger and larger bodies, which can probably only be achieved for very fluffy ice grains, otherwise fragmentation, bouncing, and radial drift limit pebble sizes to just a few mm \citep[e.g.][]{Birnstiel2012, Kataoka2013}. Or alternatively, self-gravity forces an entire cloud of pebbles to contract into planetesimals as originally pointed out by \citet{Safronov1969} and \cite{GoldreichWard1973}. \corrected{\citet{Weidenschilling1980} interjected that the pebble-gas interaction would lead to turbulent diffusion, rendering the necessary densities for gravitational instability impossible.} Meanwhile, we understand that gas turbulence \corrected{merely} regulates the onset of gravitational collapse by controlling both pebble sizes as well as the dust-to-gas ratio in the settled pebble layer around to the mid-plane of the disk \citep{Estrada2016, drazkowska2016, drazkowska2017, Lenz2019, Gerbig2019, Stammler2019}.
Magneto and hydro dynamical gas turbulence is in all cases needed to locally concentrate pebbles in a disk, be it as trapping pebbles in flow features like \corrected{in-plane horizontal vortices \citep{BargeSommeria1995}, convection like vertical cells \citep{KlahrHenning1997} and zonal flows (aka pressure bumps) \citep{Whipple1973}, as the typical dust to gas ratio in the solar nebula is too low for gravitational collapse in the presence of Kelvin Helmholtz (KHI) and streaming instability (SI) \citep{Johansen2009, Carrera2017, Gerbig2020}.  See \citet{Klahr2018} for a review on the role of turbulence and flow structures for planetesimal formation.}
Starting from a mild concentration of pebbles in a pressure bump by a factor of a few (defined as vertical integrated dust-to-gas ratio of $Z = \SigmaDust/\SigmaGas \approx$ 0.02 - 0.03), a gravitational unstable pebble cloud (local dust-to-gas ratio of $\varepsilon = \rhoDust/\rhoGas = 10 - 100$ in the solar nebula \citep{KlahrSchreiber2020a}) can then be created by turbulent clustering \citep{Cuzzi2008, Cuzzi2010, HartlepCuzzi2020}, by sedimentation and SI without \citep{Youdin2005,Johansen2009, Gerbig2020} or with additional concentration in zonal flows and vortices \citep{JohansenKlahrHenning2006, 2007Natur.448.1022J, Carrera2020}. 

What all these gravitational collapse models have in common, is that the job is not finished when the collapsing pebble cloud reaches the Hill density $\rho_{\rm Hill}$ that is when tidal forces from the central star with mass $M$ can no longer shear the pebble cloud orbiting at distance $R$ apart, 
\begin{equation}
\rho_{\rm Hill} = \frac{9}{4 \pi}\frac{M}{R^3},
\label{HillDens}
\end{equation}%
\corrected{but actually when the pebble accumulation reach the solid density of a comet or an asteroid, which is dependent on the distance to the star $10^6$ to $10^{12}$ times larger than the Hill density.}

\corrected{As mentioned above pebbles at Hill density correspond to a dust-to-gas ratio of 10 - 100 in the solar nebula at early times \citep{Lenz2020,KlahrSchreiber2020a} and even so the solids are locally dominating the dynamics, the presence of gas still can hamper the gravitational contraction as discussed by \citep{Cuzzi2008}. There are two limiting factors that determine the fate of the contracting clump.} Specifically, a pebble cloud could experience erosion by head wind or internal turbulent diffusion. \citet{Cuzzi2008} argue that turbulent diffusion is typically weaker than ram pressure from the head wind and therfore neglect the effect in their further studies \citep{Cuzzi2010, HartlepCuzzi2020}.
Yet, in order to efficiently form planetesimals at the desired small sizes of 10 - 100 km, \citet{HartlepCuzzi2020} require a significant reduction of the headwind, up to a factor of 30 in a zonal flow. As such, we argue in \cite{KlahrSchreiber2020a}, that internal diffusion cannot be neglected.
If headwind is reduced by a factor of 30, for instance in a zonal flow, then automatically diffusion will be the limiting factor.

\corrected{\citet{KlahrSchreiber2020a} compared the time scale for contraction at Hill density under self gravity with the turbulent diffusion timescale and derived a critical length of $\lcrit$, above which diffusion would we slower than contraction. Based on that paradigm they introduce a critical mass $m_{c}$, i.e.\ a sphere of radius $\lcrit$ at Hill density needed for gravity to overcome diffusion of pebbles with a size represented as Stokes number $\stokes$ for a normalized diffusivity of $\delta$:}
\begin{equation}
m_{c} = \frac{4 \pi}{3}    \lcrit^3 \rho_{\rm Hill} = \frac{1}{9} \left(\frac{\delta}{\St}\right)^{\frac{3}{2}} \left(\frac{H}{R}\right)^3 M_\sun.
\label{eq:mass}
\end{equation}
\corrected{$H/R$ is the relative pressure scale height of the protoplanetary disk, reflecting the local gas temperature.
The Stokes number is the friction time (or coupling time) of pebbles $\tau$ \citep{Weidenschilling1977} multiplied with the orbital angular velocity $\Omega$, i.e., $\St = \tauS \Omega$. It quantifies how well the particles are coupled to the gas and thus
how quickly they sediment to the midplane \citep{Dubrulle1995}, drift towards the star \citep[e.g.,][]{Nakagawa1986} and how well they drive instabilities \citep[e.g.,][]{SquireHopkins2018} and couple to turbulence \citep{2007Natur.448.1022J}.}

\corrected{For a given Stokes number, relative pressure scale height $H/R$, stellar mass $M_\sun$ and normalised strength of the SI, which means removing the actual gas disk profile from the equations, there is no explicit dependence of critical mass on distance to the star left in the expression. The reason lies in that the Hill density drops with $R^{-3}$ with distance to the star and at the same time the volume of the critical pebble cloud scales with $R^3$ (for constant $H/R$). Thus a dependence of mass on $R$ comes only from the radial profile of $H(R)/R$, $\delta(R)$ and $\stokes(R)$.} \corrected{The diffusivity $\delta$ generated by the SI appears to scale proportional to the Stokes number and inversely with the mean dust to gas ratios we have to consider here \citep{Schreiber2018,KlahrSchreiber2020a}:
\begin{equation}
\delta \approx \delta_0 \frac{10}{1 + \varepsilon_{\rm Hill}}\frac{\St}{0.1}
\end{equation}
 and $\stokes$ possibly cancels from the mass prediction (Equation\ \ref{eq:mass}). Thus ultimately the pebble to gas ratio at Hill density is left to be the dominant effect for planetesimal sizes:}
\begin{equation}
m_{c} \approx 100 \left(\frac{\delta_0}{\varepsilon_{\rm Hill}}\right)^{\frac{3}{2}} \left(\frac{H}{R}\right)^3 M_\sun.
\label{eq:mass_simple}
\end{equation}
\corrected{Further studies of pebble diffusivity in relation to disk structure and especially pebble size distribution \citep{Schaffer2018} for a range of pebble to gas ratios are therefor needed to further constrain the critical masses for planetesimal formation.}

\corrected{As in related work \citep{Nesvorny2010, WahlbergJansson2014}, we represent the mass of a pebble cloud as the equivalent (compressed) diameter, this means: if we compress a cloud with a certain mass $m_c$ from Hill to solid density $\rhoSolid \approx 1 {\rm g cm^{-3}}$, it would have a new diameter $a_c$. The actual range of planetesimal average density may fall between $\rhoSolid \approx 0.5 {\rm g cm^{-3}}$ for comets and $\rhoSolid \approx 2 {\rm g cm^{-3}}$ for some of the asteroids, but we neglect this effect for now as we do an order of magnitude estimate and only $\rhoSolid^{1/3}$ enters the expression for the compressed size.}

\corrected{In \citet{KlahrSchreiber2020a} we find equivalent sizes of gravitational unstable pebble clouds that range from $ a_c = 60 - 120$ km in a model for the early stages of the solar nebula \citep{Lenz2020}. This size range reflects the varying pebble to gas ratio at Hill density for the gas profile of the nebula and the local $H/R$, i.e.\ the temperature profile of the gas. Both $\varepsilon_{\rm Hill}$ and $H/R$ have a radial profile, yet in effect can balance out each other in terms of controlling planetesimal masses.}

The critical length $\lcrit$ in Equation\ \ref{eq:mass} is \corrected{not only} the minimum radius for a cloud of pebbles with Stokes number $\St$, at Hill density and in the presence of turbulent diffusion with strength $\delta$ acting on this length scale to collapse i.e.,
\begin{equation}
\label{eq:collapseCrit}
\lcrit = \frac{1}{3} \sqrt{\frac{\delta}{\stokes}}  H,
\end{equation}

Simultaneously, $\lcrit$ is also the scale height of the particle layer if it reaches Hill density \corrected{as its peak value}  \citep{KlahrSchreiber2020a, Gerbig2020} and, as we will show in this paper, also the characteristic radius of a Bonnor-Ebert solution for a pebble cloud with a central density of $\rhoHill$.

In \citet{KlahrSchreiber2020a}, the collapse criterion was derived for the assumption that turbulent diffusion acts isotropic in all directions. Subsequently we tested the criterion in two-dimensional simulations of the SI.
Before turning on self-gravity, we measured the diffusion for different Stokes numbers, radial pressure gradients and different box sizes, i.e.\ different mass quantities of pebbles in the simulation domain at the same dust-to-gas ratio while remaining at Hill density. With the measured diffusion, we then predicted which simulations should gravitationally collapse, and which ones should stay stable. In all cases, the prediction that the simulation domain $L$ has to be 
\konsti{larger} than $2 l_{\mathrm{c}}$ (from Equation \ref{eq:collapseCrit}) for collapse to proceed was satisfied. Roughly speaking a sphere of radius $l_c$ would have to fit into our simulation domain to allow for collapse.
 
However, all simulations \konsti{in \citet{KlahrSchreiber2020a}} were two-dimensional and radial and vertical diffusion are known to have unequal relative strengths if driven by the SI \citep{JohansenYoudin2007, Schreiber2018}. Thus, in the present paper, we study the SI in a three dimensional box, measure radial and vertical diffusion and then switch on self-gravity to check for gravitational collapse for different total mass content (pebbles plus gas) in the box.

\corrected{\citet{Li2019} present the highest resolution study in a line of papers that determine the size distribution of planetesimals formed via self gravity in the presence of streaming instability \citep{Johansen2015,Simon2016,Simon2017,Abod2018}. At least in one of their simulations using rather large pebbles $\stokes = 2$ the binned size distribution shows a maximum of objects with a diameter of 100 km. In our interpretation this turn-over in the size distribution should reflect the critical pebble mass needed for gravitational collapse in the presence of turbulent diffusion. Unfortunately, the strength of diffusion was not determined in those runs. Also despite a huge resolution in \citet{Li2019}, our boxes are still 25 higher in resolution, which maybe important to resolve the critical length scales sufficiently. In that context we can interpret our numerical experiments as a zoom in to the densest regions in \citet{Li2019} to study whether we can explain the turn over via turbulent diffusion.}

In Section \ref{sec:2}, we discuss several necessary concepts for the interpretation of our numerical simulations of self-gravitating pebble clouds: the scale height of the pebble layer subject to diffusion, and the relation and difference between Toomre and Hill stability criteria for our turbulent pebble cloud. 
In Section \ref{sec:3}, we present our numerical 3D SI simulations. We show a set of five different simulations, of which only the first is without self-gravity. In the following four simulations self-gravity is switched on and we increase the total mass of the domain for both pebbles and gas, thereby maintaining the dust-to-gas ratio and thus the potential strength of the SI.

To test our base line assumption of deriving a collapse criterion for a sphere of constant density we also interpret our simulation with a centrally peaked pebble cloud in Section \ref{subsec:3-4}. There we derived a centrally peaked Bonnor-Ebert solution for the density distribution in a pebble cloud in which now $l_{\mathrm{c}}$ determines the critical radius of a sphere with central density $\rho_{\mathrm{c}} = \rhoHill$. 
There we also discuss the effect of non-isotropic diffusion, actually creating an Bonnor-Ebert ellipsoid. 
\corrected{We summarise our results in Section \ref{sec:4}, where we also compare our results to our two-dimensional studies \citep{KlahrSchreiber2020a} and our large scale simulations on the onset of planetesimal formation \citep{Gerbig2020}}.

\correctedn{In Appendix A we reiterate the scale height of pebbles under self gravity and diffusion. In Appendix B we introduce our new concept of diffusive pressure, i.e.\ the treatment of diffusion in the momentum equation rather than in the continuity equation. Thus (angular-) momentum conservation is automatically achieved in our analysis, which we discuss in the context of secular gravitational instability. As this diffusive pressure leads to a formal speed of sound for the pebbles we discuss the implications of that concept in Appendix C.}

\begin{deluxetable*}{lll}[tb!]
\tabletypesize{\footnotesize}
\tablecaption{Used symbols and quantities:\label{tab:usedSym}}
\tablehead{\colhead{Symbol} & \colhead{Definition }& \colhead{Description}}
\startdata
$R$,$M_\odot$         &                                   & heliocentric distance, solar mass   \\
$m_c$      &                                   & critical mass of unstable pebble cloud\\
$a_c$      &           $a_c \sim m_c^{1/3}$      & equivalent compressed diameter \\
$\Omega$, $\Torb$        &     $\Torb = 2\pi/\Omega$                               & orbital frequency, orbital period \\
$t$             &                                   & time in orbital periods \\
$G, \Gmod$      &                                   & gravity constant, resp. in code units  \\
$\rho, \rho_0 $ & $\rho_0 = \rho(t=0) = <\rho>$     & local and initial (mean) pebble density \\
$\rhoGas$       &                                   & gas density                \\
$\cs,\vth,\vthoneD$      & $\cs = \vthoneD = \sqrt{\frac{\pi}{8}}\vth$ & isothermal sound speed, and 1D and 3D thermal speed (gas)       \\
$\rhoHill$      & $\rhoHill=9M/4\pi R^3$            & Hill density      \\
$\rhoSolid $    &                                   & solid body density    \\
$\aDust $       &                                   & pebble radius    \\
$\tauS$         & $\tauS = \frac{\aDust \rhoSolid}{\rhoGas \vth}$ & stopping/friction time of peppbles                                       \\
$\stokes$       & $\tauS\Omega$                     & Stokes number                                         \\
$\tauFF$        & $\tauFF = \sqrt{\frac{3 \pi}{32 G \rho}}$ & free fall time for density $\rho$                                       \\
$\tauC$         & $\tauC = \tauFF \left(1 + \frac{8 \tauFF}{3 \pi^2 \tauS}\right)$ & contraction time (incl.\ friction)                     \\
$\tau_\mathrm{t}$ && correlation time of turbulence\\
$\mathbf{u}$, $\mathbf{v}$  &                       & gas and dust velocity\\
$\scaleheight$  & $\scaleheight=\cs/\Omega$         & gas disk scale height                                 \\
$\varepsilon $     & $\varepsilon=\rho/\rhoGas$       & local dust-to-gas density ratio                             \\
$Z$             & $Z=\SigmaDust/\SigmaGas$          & dust-to-gas surface density ratio                             \\
$\varepsilon_\mathrm{max}$,$\varepsilon_0$ &              & maximum and initial dust-to-gas ratio (simulation)    \\
$\varepsilon_\mathrm{Hill}$ &                          & dust-to-gas ratio at reaching Hill density    \\
$\rho_c$        &                                   & central density in a pebble layer or a Bonnor-Ebert sphere \\
$\rho_\circ$    &                                   & density at the surface of a Bonnor-Ebert sphere \\
$f$             & $f = \frac{\rho_0}{\rhoHill}$     & initial pebble density in simulation  \\
$L$             & $L = 0.001 H$                     & simulation domain size                    \\
$\nu,\alpha$    & $\nu = \alpha \cs \scaleheight$   & global viscosity / diffusion coefficient \\
$D$,$\delta_{(x,z)}$    & $D =\delta  \cs \scaleheight$     & local / small scale (anisotropic) diffusion coefficient \\
$h_p$           &$h_p =  \sqrt{\frac{\delta}{\stokes}} H$ & Pebble Scale height without self gravity \\
$a$             & $a = \sqrt{\frac{\delta}{\stokes}}c_s$ & pseudo sound speed of pebbles under diffusion  \\
$P$             & $ P = a^2 \rho $ & pseudo pressure for pebbles under diffusion \\
$\lcrit, \lcritx, \lcritz$ &$\lcrit = \frac{1}{3} \sqrt{\frac{\delta}{\stokes}} H$    & critical length / Scale height for $\rho_c = \rhoHill$\\
$\hlcrit$     &$l_c = \frac{1}{3} \sqrt{\frac{\delta}{\stokes}}\sqrt{\frac{\rho_{\rm Hill}}{\rho_c}} H$    & same for $\rho_c > \rhoHill$\\
$Q_g, Q_p$      &   $Q_p = \sqrt{\frac{\delta}{\stokes}} \frac{Q_g}{Z}$ & Toomre parameter for gas and pebbles\\
$\lambda_{\rm fgm, min, max}$ &                     & fastest smallest and largest Toomre wavelength\\
$\omega$,$\gamma$ & & frequency, growthrates of plane waves\\
$\lambda_{\rm Jeans}$ &                & "Jeans" wavelength \\
$M_{\rm Jeans}$ &                & mass of marginally stable Bonnor-Ebert sphere\\
$\eta, \beta$   &  & pressure gradient parameters 
\enddata
\end{deluxetable*}

We will follow the notation in Tab. \ref{tab:usedSym} throughout this paper.

\section{Self-gravity of particle layers}
\label{sec:2}
\citet{Safronov1969} and \cite{GoldreichWard1973} considered the gravitational stability of a particle layer in the solar nebula for the case that gas can be ignored and derived dispersion relations and probable planetesimal masses to result from gravitational fragmentation. Yet, as shown by \citet{Weidenschilling1980} the interaction with the gas cannot be neglected as it can drive turbulence. Turbulent diffusion limits sedimentation and thus appears to prevent the necessary concentration of pebbles for self-gravity to become important. But this is only true if one considers turbulence to be a strictly diffusive process. As we know today, turbulence also concentrates material, either as part of a particle-gas instability \citep{Youdin2005}, via turbulent clustering \citep{Cuzzi2008}, or through trapping in non-laminar flow features \citep{Whipple1973,BargeSommeria1995,KlahrHenning1997}.

\citet{Sekiya1983} included gas for the gravitational stability of the particle layer, but he considered a closely coupled dust and gas system, effectively $\stokes \ll 1$. Finite coupling times were introduced to study a secular gravitational instability \citep{Ward1976,Ward2000,Coradini1981, Youdin2005} in which particle rings contract radially, thereby losing excess angular momentum due to friction with the gas \citep{ChiangYoudin2010} . \corrected{In those studies one considers the motion of the pebble swarm at its rate of terminal velocity with respect to the gas, as a consequence of the rotational profile of the nebula \citep{Sekiya1983} and the mutual gravity of the pebbles.} \corrected{Diffusion of pebbles via turbulence has also been added to these studies \citep{Youdin2011} in explicitly adding a diffusion term to the mass transport of pebbles. Recently \citet{Tominaga2019} showed that the diffusive pebble flux should also be treated in the momentum equation to ensure angular momentum transport.}

\corrected{What we do differently in our stability analysis, is to treat the diffusion of particles via turbulent mixing in the momentum equation instead of the continuity equation. As derived in appendix \ref{sec:B} we 
define the pebble velocity in the continuity equation as the sum of advective and diffusive flux. Redefining the momentum equation to this new pebble velocity introduces a source term for the momentum equation that looks formally like a gradient in pebble pressure $P$.} 
\begin{equation}
    \partial_t \mathbf{v} \rho = - \frac{D}{\tau} \mathbf{\nabla} \rho := -\mathbf{\nabla} P.
    \label{eq:Danda}
\end{equation}
\corrected{The formal speed of sound $a$ related to this pressure gradient is diffusivity $D$ divided by the stopping time $\tau$}
\begin{equation}
    a^2 = \frac{D}{\tauS}.
    \label{eq:A2anda}
\end{equation}
\corrected{and not the actual r.m.s.\ velocity of the particles $v_\mathrm{r.m.s.}$, which is proportional to diffusivity divided by the correlation time of turbulence \citep{YoudinLithwick2007}}
\begin{equation}
    v_\mathrm{r.m.s.}^2 = \frac{D}{\tau_\mathrm{t}}.
    \label{eq:YLrms}
\end{equation}
\corrected{Our derivation uses a balance between diffusion and sedimentation via the momentum equation. This is common practice, for example, when calculating the scale height of the particle layer in the midplane of a turbulent disk.}
\corrected{We assume the gas to be turbulent, yet incompressible, which holds even during the gravitational contraction of the pebble cloud as shown in \citep{KlahrSchreiber2020a}. The particles can move with respect to the local gas velocity fluctuations, which on average are zero. Thus, in first order approximation particles have to sediment and contract with respect to the gas at rest and get diffused by turbulence, which for the evolution of the pebble distribution $\rho$ acts as a gradient of the pebble pressure $P = a
^2 \rho$, with the speed of sound of the pebbles $a$ being a fraction of the gas speed of sound $\cs$.}
\begin{equation}
    a = \sqrt{\frac{\delta}{\stokes + \delta}} \cs \approx \sqrt{\frac{\delta}{\stokes}} \cs.
    \label{eq:asound}
\end{equation}
\corrected{Such a relation between friction $\tauS$, diffusion $D$ and a "thermal" velocity $a$ is not new. It is the same derivation of diffusivity $D_\mathrm{B}$ for a particle of mass $m$  under Brownian motion at temperature $T$ and with the friction parameter $\mathrm{f} = m / \tauS$  found by \citet{Einstein1905}, also based on an equilibrium of diffusion and sedimentation under gravity:}
\begin{equation}
    D_\mathrm{B} = \frac{\mathrm{k} T}{\mathrm{f}} = \tauS \vthoneD^2.
    \label{eq:Einstein}
\end{equation}
\corrected{$\mathrm{k}$ is here the Boltzmann constant and $\vthoneD = \sqrt{\frac{k T}{m}}$ the one dimensional thermal velocity of the particle. For a gas the one dimensional thermal velocity is also the isothermal speed of sound $\vthoneD = \cs$. We thus can associate the turbulent gas in the astrophysical environment with the heat bath that drives Brownian motion.} 

\correctedn{Using expression \ref{eq:YLrms} we can now also relate the r.m.s.\ speed of pebbles $v_\mathrm{r.m.s.}$ with the pseudo sound speed $a^2$ of pebbles as}
\begin{equation}
   a^2 = v_\mathrm{r.m.s.}^2 \frac{\tau_\mathrm{t}}{\tau}.
   \label{eq:a_v_relation}
\end{equation}
\correctedn{For pebbles of $ \frac{\tau_\mathrm{t}}{\tau} = 1$ the r.m.s.\ speed and pebble sound speed are then identical. But for larger pebbles the sound speed decreases and for smaller pebbles it increases. The latter case is then usually limited by the compressibility of the gas and set to the speed of sound as done by \citet{Dubrulle1995}.
}

\corrected{Equation \ref{eq:asound} can easily be transformed to an equation of the thickness of the pebble layer $h_p$ in \citet{Dubrulle1995} by dividing both sides by $\Omega$ and with $c_s = H \Omega$} it follows:
\begin{equation}
    h_\mathrm{p} := \frac{a}{\Omega} = \sqrt{\frac{\delta}{\stokes + \delta}} H,
    \label{eq:Dubrulle}
\end{equation}
\corrected{which will be a handy expression for this paper.}

\correctedn{In the appendix \ref{sec:C} we show that in fact $a$ is the speed of sound of wave like perturbations of pebbles under diffusion, but for physical realistic wave numbers, those waves are critically damped in less than one oscillation period. Only for nonphysical short wave lengths, where the diffusion description would break down, one can mathematically derive oscillatory solutions.}

While SI is just one possible origin of local gas turbulence, it is the easiest to be studied in small boxes at a fraction of the gas pressure scale height and requires fewer assumptions compared to introducing additional external $\alpha$ turbulence stemming from large scales as recently done by \citet{Gole2020}. Additionally, SI dominates on the collapse scales of pebble clouds \citep{KlahrSchreiber2020a} and is therefore ideally suited for our investigation. \corrected{So we distinguish between large scale turbulence $\alpha$ introduced by \citet{ShakuraSunyaev1973} to parametrise angular momentum transport and that may stem from magneto hydro instabilities \citep{BalbusHawley1998} or hydro dynamic instabilities \citep{KlahrBodenheimer2003,Nelson2013,Marcus2016}, which seem to be relevant in protoplanetary disks \citep{Pfeil2019}. Even so $\alpha$ may also have a non-turbulent wind component \citep{Bai2013,Bethune2017}, for lack of better knowledge $\alpha$ is also assumed to drive global diffusion and pebble collisions that determine the conditions for planetesimal formation in terms of dust-to-gas ratio and Stokes number from the large scales \citep{Schaffer2018,Gerbig2020}. And on the other hand we define $\delta$ as the local small scale diffusivity on the scales of pebble cloud collapse. At large scales $L \sim H$ the assumed $\alpha$ is typically orders of magnitude larger than $\delta$, but once $\alpha$ is cascaded down \citep{Kolmogorov1941} to the scales relevant to form a planetesimal of less than 100 km from a pebble cloud at Hill density $L \sim 0.001 H$, then $\delta$ is predominantly produced locally by SI and other resonant drag instabilities \citep{Squire2018}.}

As pointed out by \citet{Gerbig2020}, even in the absence of global $\alpha$ turbulence the SI is not the only effect setting the vertical scale of the particle mid-plane, but at the expected high dust to gas ratios and small scales, on a first guess the most important one.
Nevertheless, we will see in Section 3.2 that as soon as self gravity is included, even in our non-stratified disk the conditions for Kelvin-Helmoltz instability are given, enhancing the strength of particle diffusion. As such, we will first show how the vertical scale height of the pebble layer in a turbulent disk is modified with the inclusion of self-gravity.

\subsection{Dust scale height at Hill density}

\begin{figure*}
\gridline{
    \fig{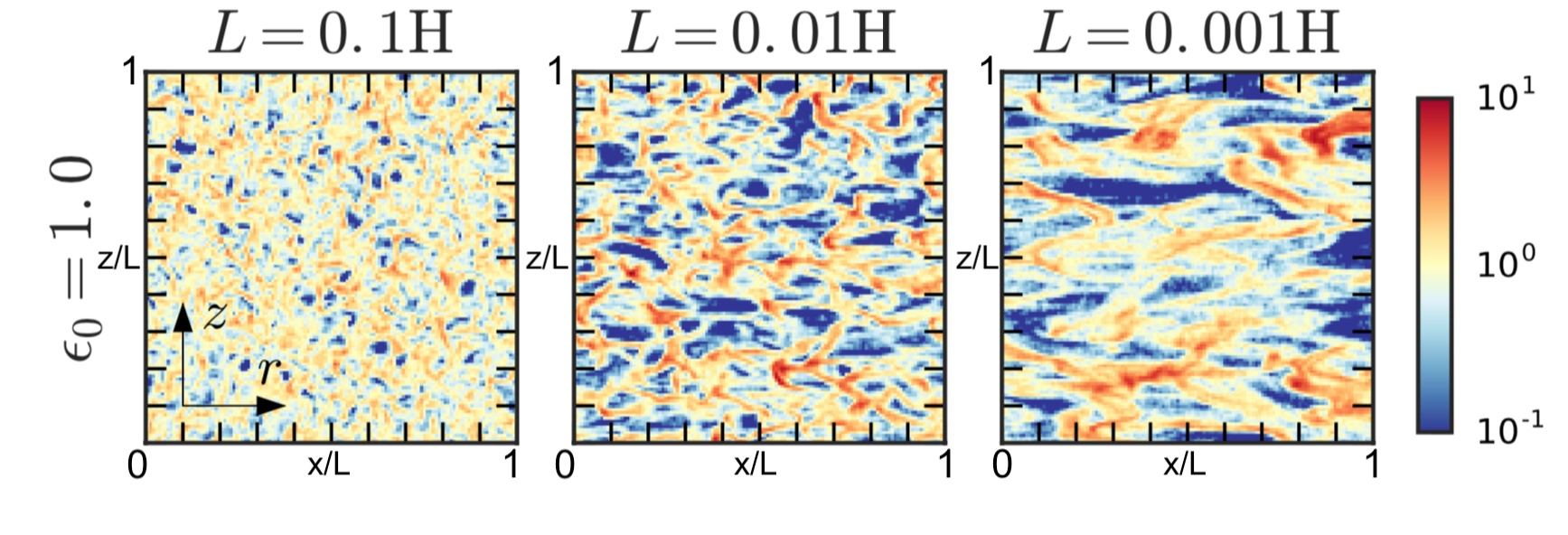}{0.9\textwidth}{(a): Radial-vertical slice for $\varepsilon/\varepsilon_0$.}
}
\gridline{\fig{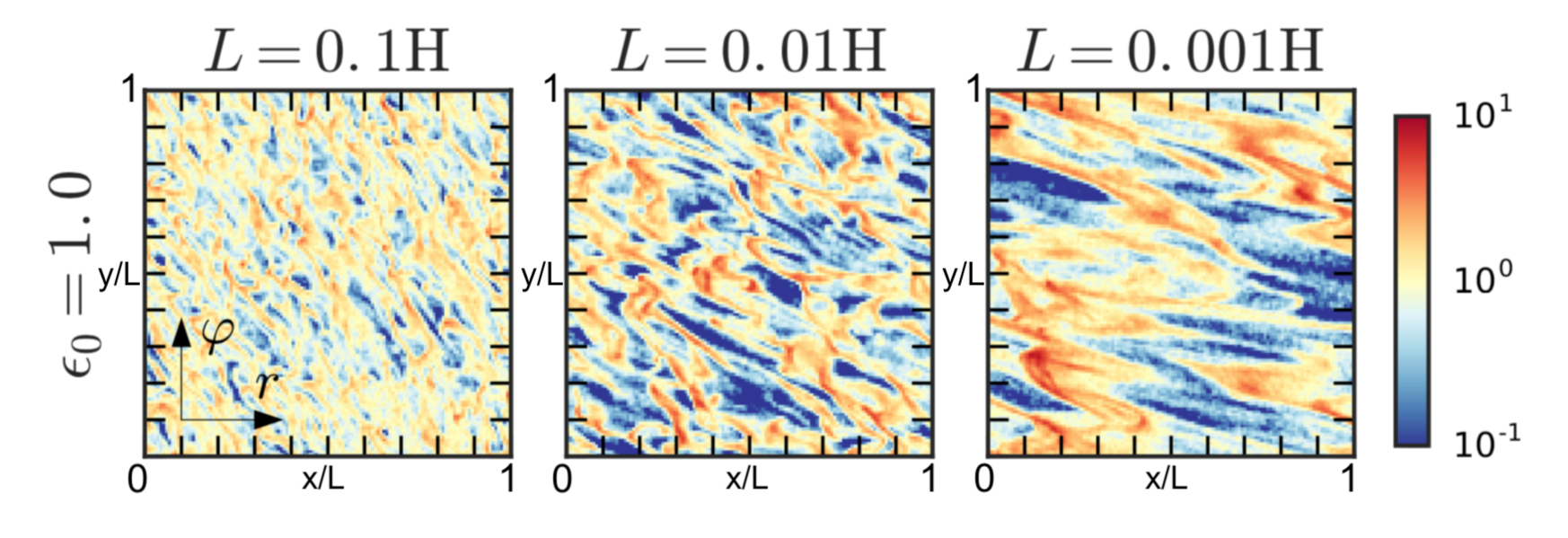}{0.9\textwidth}{(b): Radial-azimuthal slice for $\varepsilon/\varepsilon_0$.}
}
  \caption{Simulation end-states of our 3D SI study without self gravity. Colour scale represents dust to gas ratio $\varepsilon/\varepsilon_0$ between 0.1 and 10.0. We compare 3 different box sizes, but always the same physical parameters in terms of particle size and average dust to gas ratio. The upper row is a slice in the radial vertical direction and the lower row in the radial azimuthal direction. The left column with $L = 0.1 H = 20 \eta R$ is an almost copy of model AB-3D ($L = 40 \eta R$) in \citep{JohansenYoudin2007} and shows the same saturated state in their Figure 9. For our collapse study we increased the resolution in steps of factor 10 while reducing the box size. Therefor we see basically a zoom in to smaller scales of SI. As smaller scales are less unstable the overall strength of SI was getting weaker, but simultaneously we were able to find the lower cut off of instability, as the smallest strictures are clearly resolved. We chose the $L=0.001 = 0.02 \eta R$ simulation for our self-gravity study, because based on the measured diffusion, the critical length could be larger than the box size, preventing collapse.}
  \label{fig:3d_prep_simulations}
\end{figure*}

In our previous numerical experiments \citep{KlahrSchreiber2020a}, we performed two-dimensional vertically integrated simulations of SI and self-gravity. 
This means that the simulations were effectively 2D, i.e.\ the third dimension is entirely in one cell and thus vertically integrated by default. Thus, we did not have to consider sedimentation of the particles. In the present three-dimensional work, we do not have vertical stellar gravity either, because for us it is sufficient to study streaming instability (SI) only and also do not cover sufficient height of the disk to induce Kelvin Helmholtz instability (KHI): the study of KHI modes driven by the sedimentation of dust \citep{Weidenschilling1980} demands larger boxes of about $L = 0.4H$ as seen in e.g.\ \citet{Gerbig2020}. \corrected{In that paper it was the goal to understand the needed dust enhancement $Z$ to overcome KHI to create the necessary dense pebble layer to trigger streaming instability and self-gravity for planetesimal formation. It was the question whether it is possible to form any planetesimal independent of size. In contrast, for the present paper we 
assume that we are already in the situation that planetesimals can principally form, as Hill density is already reached, but we ask how big a pebble cloud has to be in order collapse. To answer this question,} we want to identify the smallest possible box at Hill density or more precisely, the smallest necessary mass in a small box that can undergo gravitational collapse. Thus the box in our numerical experiments is only $L_x = L_y = L_z = 0.001 H$ in size. We have chosen that size because for the Stokes number we picked and the strength of SI \corrected{in terms of measured diffusivity} we found, the expected critical length-scales $l_c$ should also be on the order of $0.001 H$ (see Figure \ \ref{fig:3d_prep_simulations}).
\corrected{In \citet{Schreiber2018} we experimented with the influence of Stokes number, average dust-to-gas ratio and box size on the strength of streaming instability in terms of pebble density fluctuations and particle diffusion. We found that SI becomes weaker as soon as the fastest growing modes are not fitting into the box anymore, yet SI will not die out and still drive significant diffusion controlling the onset of gravitational collapse at high dust to gas ratios. Other work usually does not consider such high dust to gas ratios or small boxes, yet for the parameters where we approach the simulations of \citet{JohansenYoudin2007} we find an agreement in the measured SI properties in terms of r.m.s. velocities, diffusion and particle concentration.}

Without self-gravity
\corrected{
the thickness of the pebble layer around the midplane under turbulence for $\stokes > \delta$ is (see Equation \ref{eq:Dubrulle})}
\begin{equation}
h_\mathrm{p} = \sqrt{\frac{\delta}{St}} H.
\end{equation}
However, upon reaching Hill density at the midplane, the vertical acceleration from self-gravity $g_z = -\partial_z \Phi_{\rm dust}$ acting on the dust is nine times stronger than the vertical component of stellar gravity as seen from the Poisson equation 
\begin{equation}
\nabla^2 \Phi_{\rm Hill} = 4 \pi G \rho_{\rm Hill} = 9 \Omega^2,
\end{equation}
where we used the definition of the Hill density in Equation~\ref{HillDens}.
Thus, around the mid-plane gravitational acceleration is $g_z = - 9 \Omega^2 z$ in comparison to stellar gravity $g_\odot = -\Omega^2 z$ and we can neglect the latter.
As shown in appendix \ref{sec:B} this leads to a new pebble scale height at Hill density of
\begin{equation}
\lcrit = \frac{1}{3} \sqrt{\frac{\delta}{St}} H
\label{eq:firstlc}
\end{equation}
and expanding this to even larger peak densities $\rho_c$ (see Equation\ \ref{eq:firsthlc}) gives:
\begin{equation}
\hlcrit = \sqrt{\frac{\rho_{\rm Hill}}{\rho_c}} l_c.
\label{eq:firsthlca}
\end{equation}
In \citet{KlahrSchreiber2020a} we show that the vertical distribution of pebbles is actually a hyperbolic function, yet sufficiently similar to that of a Gaussian of the same width, for small values of z, i.e.\ $z < \hlcrit$.
But, if you integrate $\rho$ vertically from $-\infty$ to $+\infty$ for a Gaussian you receive $\Sigma_\mathrm{Gauss} = \sqrt{2 \pi} \hlcrit$, whereas for the hyperbolic function it is $\Sigma = 2 \sqrt{2} \hlcrit$, which is what we use for our further analysis.
The average or initial dust density $\rho_0$ that we choose for our computational domain is defined by multiples $f$ of the Hill density $\rho_0 = f \rhoHill$,
thus the column density is always $\Sigma = L f \rhoHill$. As long as $\hlcrit$ is sufficiently smaller than the box height $L/2$, 
we can use 
\begin{equation}
\Sigma = 2 \hlcrit \sqrt{2} \rho_c  = f L \rho_{\rm Hill}.
\label{eq:2hat}
\end{equation}
Thus, we find by eliminating $\rho_{\rm Hill}/\rho_c$ via Equation\ \ref{eq:firstlc}
an expression to determine the vertical diffusivity $\delta$ in our numerical experiments (with fixed $L,f,\stokes$ and $H$) as a function of the measured pebble scale height $\hlcrit$: 
\begin{equation}
\delta = \frac{9}{2 \sqrt{2}} f \stokes \frac{L \hlcrit}{H^2}.
\label{eq:firstlc3}
\end{equation}
Note that in case of self gravity the diffusivity is proportional to the dust scale height $\hlcrit$ whereas in the case of no self-gravity diffusivity is proportional to the square of the dust scale height $h_\mathrm{p}^2$ (see Equation \ref{eq:Dubrulle}).

\subsection{Toomre stability}

As explained in \citet{KlahrSchreiber2020a}, the criterion of $\lcrit < \onehalf L$ \corrected{(a sphere of radius $\lcrit$ has to fit into the simulation box with dimensions $L$)} for gravitational collapse describes the stability of a local non-linear density fluctuation. In contrast, the Toomre stability criterion \citep{Toomre1964} applies to the linear growth of infinitesimal perturbations in surface density, so it is worthwhile to reconcile the relation between the two criteria here.

\corrected{The Toomre analysis for planetesimal formation in \citet{Safronov1969} and \citet{GoldreichWard1973} considers a gas free system with the random motions of (almost collision free) particles providing a pressure counteracting gravity. The root mean square of these random particle velocities then defines the "sound speed" of the pebbles. This leads to the same approach as when considering the gravitational stability of a gas disk with thermal pressure and the speed of sound of the gas \citep{BinneyTremaine2008}.} 

\corrected{But note, that if we now also derive a "sound speed" for pebbles diffused by turbulence, then this is not simply the velocity dispersion of the pebbles.}
\corrected{Our particle speed of sound represents the resistance of pebbles clouds against compression by gravity. For instance in case of negligible turbulent diffusion $\delta \rightarrow 0$, but having a Stokes number still smaller than $\delta$, the "particle speed of sound" approaches the speed of sound of the gas (see Equation\ \ref{eq:asound}), whereas the velocity dispersion approaches zero (See Equation\ \ref{eq:YLrms})}. 

\corrected{This also means, that one cannot use the measured r.m.s.\ velocities of the pebbles in our simulation for the Toomre analysis, but one needs the actual diffusivity on the scales of accumulations. Therefore we measure the scale height of pebbles in the disk (see Equation\ \ref{eq:firstlc3}) and track the diffusion of individual pebbles \citep{KlahrSchreiber2020a}.}

We write the momentum equation for our pebble-gas in a classical fashion \citep[see e.g.,][]{ChiangYoudin2010}, but instead of thermal pressure or a dispersion velocity, turbulent diffusion acts as the stabilizing agent.
As a result, the momentum flux by pressure \corrected{for an ideal gas} $ - c_s^2 \nabla \rho $ is
replaced by the \corrected{the "diffusion pressure" for closely coupled particles } $- a^2 \nabla \rho = - \frac{D}{\tau}\nabla \rho$, \corrected{which means that the diffusive flux is generated in the momentum equation and not added to the continuity equation (see our derivation in appendix \ref{sec:A}). Thus the pebble velocity $\mathbf{v}$ already contains the diffusive flux and one avoids the problem that that neglecting the diffusive flux in the momentum equation can lead to the violation of angular momentum conservation \citet{Tominaga2019}.}
Thus, we adopt an only slightly modified set of equations to describing the hydrodynamic behaviour of pebbles under self-gravity in comparison to classical work \citep{Safronov1969,GoldreichWard1973}. \corrected{The background state for our Toomre analysis is a constant surface density of pebbles and gas and Keplerian shear. But note that we neglect azimuthal friction with the gas, which is the driver of the secular gravitational instability \citep{Youdin2011}, because for the high dust to gas ratios we consider, such a damping seems inefficient. As discussed in appendix \ref{sec:B} the radial friction could be included, but it only slows down radial contraction, but not the resulting Toomre stability criterion itself. 
We linearize the equations  for continuity and momentum around the background state as outlined in Chapter 6 of \citet{BinneyTremaine2008}}:
\begin{eqnarray}
\label{eq:BTrho}
\frac{1}{\Sigma}\partial_t \Sigma' + \partial_r  v_r' &=& 0,\\
\partial_t v_r' - 2 \Omega v_\phi' &=& -\frac{1}{\Sigma} \frac{D}{\tau} \partial_r \Sigma' - \partial_r \Phi',\label{eq:BTvr}\\
\partial_t v_\phi' + \frac{\kappa^2}{2 \Omega} v_r' &=& 0,\label{eq:BTvphi}
\end{eqnarray}
and the usual $\Phi' = 2 \pi G \Sigma' / |k|$ for the perturbed potential of a razor thin disk \citet{BinneyTremaine2008}, where
the delta function for the vertical density stratification is implemented as
\begin{equation}
    \rho = \frac{k \Sigma}{2} e^{- |k| z}.
    \label{eq:razor_rho}
\end{equation}
We adopt Wentzel-Kramers-Brillouin (WKB) waves such that perturbations scale as 
\begin{equation}
    \Sigma' = \Sigma_a e^{- i (k r - \omega t)},
\end{equation}
The dispersion relation is identical to the one given in \citet{GoldreichWard1973}, except that we replace the random motions $c$ in their equation by \corrected{our pseudo speed of sound for the particles $a$ which, to stress this once more, is not the r.m.s.\ velocity of particles in the flow. 
This pseudo speed of sound} $a \equiv c_s \sqrt{\delta_x/\stokes}$ reflects the pressure like resistance against compression, generated by diffusion. 

The dispersion relation is then:
\begin{equation}
\label{eq:dispersion_relation}
    \omega^2 = a^2 k^2 - 2 \pi G \Sigma |k| + \kappa^2,
\end{equation}
where the epicyclic frequency $\kappa^2$ is the Keplerian frequency for a Keplerian rotation profile $\kappa^2 = \Omega^2$.
We find the Toomre value for this system to be
\begin{equation}
\label{eq:particle_toomre_q}
    Q_p = \sqrt{\frac{\delta_x}{St}} \frac{c_s \Omega}{\pi G \Sigma_{\rm pebble}} = \sqrt{\frac{\delta_x}{St}} \frac{Q_g}{Z},
\end{equation}
allowing us to
directly compare the particle to the gas Toomre value $Q_g$ by using the 
metallicity $Z = \Sigma/\Sigma_{\rm gas}$. 
For $Q_p < 1$ the system is linearly unstable to perturbations.

Note, that the metallicity in the context of Equation~\ref{eq:particle_toomre_q} quantifies the 
particle concentration in the back ground state. 
A local concentration of pebbles and thus a local increase of $Z$ on scales $< 1/k$ can still be fragmenting as shown by \citet{Johansen2009} and more recently also discussed in \citet{Gerbig2020}, but this process is then not triggered by the linear gravitational instability. Still it is interesting to note that this non-linear triggered collapse will occur when the $Q_p$ calculated for a local metallicity enhancement falls below $2/3$, which is the \corrected{definition of the collapse criterion $1 > \tilde Q_p = 3/2 Q_p$ in \citet{Gerbig2020}. Note that $\tilde Q_p$ is based on the assumption of isotropic diffusion and therefore uses the vertical pebble scale height to estimate diffusivity. Yet the Toomre value $Q_p$ is independent on vertical diffusion as we have demonstrated.} 

We deem it instructive to investigate the Toomre parameter
for when the particle mid-plane reaches Hill density $\rho_c = \rhoHill$. \corrected{Yet, due to the fact that
the Toomre ansatz assumes $\Sigma = 2 \rho_c/k$, makes $Q_p$ a function of wave number for a given volume density in the midplane, this question is not straight-forward to answer.}

We begin by determining what density at the midplane satisfies $Q_p = 1$. Setting $Q_p = 1$ also defines a unique unstable wave-length of the fastest growing mode with wave number $k_{\rm fgm} = \sqrt{St/\delta} H^{-1}$. Then, via Equation\ \ref{eq:razor_rho}, we can determine $\Sigma(Q_p = 1) = 2 \rho_{\rm Toomre} H \sqrt{\delta/St}$ and  define the  Toomre density $\rho_{\rm Toomre}$ for isotropic diffusion $\delta = \delta_x = \delta_z$:
\begin{equation}
    \rho_{\rm Toomre} = \sqrt{\frac{\delta}{St}}\frac{ c_s \Omega }{\pi G 2 H \sqrt{\delta/St}} = \frac{1}{2 \pi} \frac{M_\odot}{R^3},
\end{equation}
which is $4.5$ times lower than the Hill density\footnote{This Toomre density also applies to a gas disk, as thermal pressure is generally isotropic.}.
This comes as no surprise as the Toomre criterion is for axis-symmetric modes which are not subject to tidal shear. Therefore, a local (non-axis-symmetric) particle cloud with $\rho_\mathrm{c} = \rho_\mathrm{Toomre}$ is not stable against tidal gravity and will be ripped apart. Conversely the Toomre parameter of a disk with isotropic diffusion at Hill density would fall smaller than 1 \citep{Gerbig2020}. 

On the other hand, if the dust layer is vertically much thinner than the radial unstable wavelength, because vertical diffusion is much weaker than radial diffusion ($\delta_z < \delta_x$), then Hill density in the midplane can be compatible with Toomre values larger than one. From Equation\ \ref{eq:firstlc}, we know that the vertical thickness of the relevant particle layer is $\lcrit$.
Due to potentially different radial
and vertical diffusivities ($\delta_{z} \ne \delta_{x}$), we define a new $\lcritz = \frac{1}{3}
\sqrt{\delta_z/\St} H$. 
Thus, $\Sigma = 2 \sqrt{2} \rho_{\rm Hill} \lcritz $ (see Equation\ \ref{eq:2hat}) and the Toomre value would then be
\begin{equation}
    Q_{\rm Hill} = \sqrt{\frac{\delta_x}{St}} \frac{c_s \Omega}{\pi G \rho_{\rm Hill} 2 \sqrt{2} \lcritz } = \frac{\sqrt{2}}{3}\sqrt{\frac{\delta_x}{\delta_z}}.
\end{equation}
Thus as long as radial diffusion is sufficiently larger than vertical diffusion, which seems to be case for all known studied configurations so far, then one could even globally ($L > 1 / k$) reach the Hill density in the midplane and still not be linear unstable in the Toomre fashion.

In our numerical experiments, we set the dust mass in our domain $M_\mathrm{box} = f L^3 \rhoHill$ to a fixed value, which defines the pebble surface density as $\Sigma = f L_z \rhoHill$.
As a result, the Toomre value in the simulation is given via the height of the simulation box $L_z = L$ and the mean density of pebbles expressed in multiples $f$ of the Hill density
\begin{equation}
    Q(f) = \frac{4}{9} \sqrt{\frac{\delta_x}{St}} \frac{H}{L_z}\frac{1}{f},
  \label{eq:QF}
\end{equation}
which is independent of the strength of vertical diffusion.
For the sake of completeness we calculate the range of unstable wavelengths from the Toomre parameter, which shows the fastest growing mode with $\lambda_{\rm fgm}$ to be
\begin{equation}
    \lambda_{\rm fgm} = 2 \pi \sqrt{\frac{\delta_x}{St}} H Q_p = 6 \pi Q_p  \lcrit,
\end{equation}
which is a wavelength considerably larger than the critical length for collapse $\lcrit$.
For our numerical experiments, once we have determined the strength of radial diffusion, and find Toomre values lower than unity, we can also determine a fastest growing mode for our simulations to be 
\begin{equation}
    \lambda_{\rm fgm} = \frac{8 \pi}{9} \frac{\delta_x}{St} \frac{H^2}{f L_z} = 8 \pi \frac{\lcrit^2}{f L_z}.
    \label{eq:lambda_fgm}
\end{equation}
For any Toomre parameter, when $Q < 1$ there exists a largest and smallest unstable wavelength, where large modes are stabilised by the Coriolis forces, i.e.\ $\kappa^2$
\begin{equation}
    \lambda_{\rm min, max} =  \frac{\lambda_{\rm fgm}}{1 \pm \sqrt{1 - Q_p^2}}.
\end{equation}
Interestingly the Toomre wavelengths are proportional to $\lcrit^2$, where as the  Jeans length for gravitational instability, i.e.\ in the absence of the Coriolis term
is linear in $\lcrit$
\begin{equation}
    \lambda_{\rm Jeans} = 2 \pi \lcrit,
     \label{eq:Jeans}
\end{equation}
as discussed in \citet{KlahrSchreiber2020a}. With all relevant length scales established as a function of the turbulent diffusivity, Stokes number, and the pebble load of our disk, we have all the tools at hand to interpret the following simulations.

\section{Numerical Experiments}
\label{sec:3}

\corrected{In \citet{KlahrSchreiber2020a} we used in total 15 different two-dimensional simulation set ups to study two different Stokes numbers and a range of total box sizes plus some additional runs for a different radial pressure gradient and a different initial dust-to-gas ratios. All simulations confirmed our criterion $L > 2 l_c$ respectively $m > m_c$ to form planetesimals. Such an extended parameter study is currently not possible for three-dimensional simulations. We therefore studied a range of three-dimensional SI simulations (resembling model {AB-3D} with $\stokes = 0.1$ and initial dust to gas of $\varepsilon_0 = 1$ from \citet{JohansenYoudin2007}) with decreasing box sizes $L$ (see Figure \ref{fig:3d_prep_simulations}) \citep{SchreiberThesis} and pick a particular model for which the measured diffusivity would lead to critical length scale on the order of that $L = 0.001 H$, this box is 200 times smaller than the simulation in \citet{JohansenYoudin2007}. The computational cost at the resulting resolution is immense, thus when we found that our initial model for $\rho_0 = \rhoHill$ did not collapse, we did not set up a new simulation with a bigger box, but gradually increased the total mass content in the simulations, thus not changing the dust-to-gas ratio and SI, but only the effect of self gravity.} 

\corrected{In that sense the three-dimensional simulations in the present paper are testing our collapse criterion $m > m_c$ by variation of a different parameter than in \citep{KlahrSchreiber2020a}. But in both cases, whether we change box size $L$ and keep $\rho_0 = \rhoHill$ constant for the two-dimensional case, or we keep the box size $L$ constant and increase $\rho_0 = f \rhoHill$ with $f = 2, 4$ and $8$, we effectively change the total mass of pebbles in the box, until we find the simulation to produce planetesimals. For the purpose of testing our collapse criterion, we therefor extended the definition of critical length $\lcrit$, which was originally in \citet{KlahrSchreiber2020a} only for the pebbles as Hill density, to $\hlcrit = \lcrit f^{-\frac{1}{2}} $ as a function of the increased pebble density $f$. 
Nevertheless, any size estimates for planetesimals would still be based on the original condition using $\lcrit$ with $f=1$ based on large scale SI simulation (see Equation\ \ref{eq:mass}). The modification for our numerical experiments, is justified because the general criterion $\lcrit < \onehalf L$ has to be valid for arbitrary combinations of diffusivity and pebble mass in the box. We will compare the results from this three-dimensional study with the original two-dimensional study \citep{KlahrSchreiber2020a} in Section \ref{sec:4}.}

All our past simulations \citep{Schreiber2018,KlahrSchreiber2020a} as well as those in the present paper have been performed with the Pencil Code
\citep{Brandenburg2001}, which solves for the gas density $\rho_\mathrm{g}$ with a finite difference version of the following set of equations in the shearing sheet approximation
\begin{equation}
    \frac{\partial \rho_\mathrm{g}}{\partial t} + \nabla \cdot \left(\rho_\mathrm{g} \mathbf{u}\right) + u_{0,y} \frac{\partial \rho_\mathrm{g}}{\partial y} = f_\mathrm{D}\left(\rho_\mathrm{g}\right),
\end{equation}
where $f_\mathrm{D}\left(\rho_\mathrm{g}\right)$ is a hyper-diffusivity term to stabilise the scheme. Vectors are denoted as bold figures. $\mathbf{e_x}$ and $\mathbf{e_y}$ are the unit vectors. Gas velocities $\mathbf{u}$ are solved relative to the \konsti{unperturbed local azimuthal velocity} $u_{0,y}$ \konsti{$= -q\Omega x$}, \corrected{with $q = 1.5$ for the Keplerian profile,} via the equation of motion
\begin{eqnarray}
\label{eq:gas_euler_code}
     \frac{\partial \mathbf{u}}{\partial t}  + \left(\mathbf{u}\cdot\nabla\right)\mathbf{u} +  u_{0,y}\frac{\partial \mathbf{u}}{\partial y} = -c_\mathrm{s}^2 \nabla \ln \rho_\mathrm{g} + \Omega h \beta \mathbf{e_{x}}\\
     +  \left(2\Omega u_y \mathbf{e_{x}} - \frac{1}{2}\Omega u_x \mathbf{e_{y}} \right)  - \varepsilon \frac{ \mathbf{u - v}}{\tau} + f_\nu \left(\mathbf{u},\rho_\mathrm{g}\right).\nonumber
\end{eqnarray}
Our simulation is isothermal and we use a fixed speed of sound $c_s$ and $\beta$ denotes the radial pressure gradient, see below.
Note that in contrast to \citet{Gerbig2020}, there is no vertical gravity included $- \Omega z \mathbf{z}$ as discussed before.
$\Omega h \beta \mathbf{{x}}$ represents the effect of the global pressure gradient in the disk 
\konsti{which drives the relative velocity between particles and gas \citep{Nakagawa1986}, and as such leads to radial drift and ultimately to drag instabilities like the SI.}
Particles are treated as Lagrangian tracer. Their positions $\mathbf{x}$ and velocities $\mathbf{v}$ in the shear frame are governed by
\begin{eqnarray}
\frac{\partial \mathbf{x}}{\partial t} = - q \Omega \mathbf{x} \mathbf{e_{y}} + \mathbf{v}
\end{eqnarray}
and
\begin{eqnarray}
\frac{\partial \mathbf{v}}{\partial t} = \left(2\Omega v_y \mathbf{e_{x}} - \frac{1}{2}\Omega v_x \mathbf{e_{y}}\right) - \frac{\mathbf{v - u}}{\tau}
\end{eqnarray}
with the coupling term  $- (\mathbf{v - u})/\tau$ transferring momentum between dust and gas. For additional technical features we refer to \citet{Gerbig2020}.

For our experiment, we choose a domain size \konsti{of} $L_x=L_y=L_Z={0.001}{H}$, and \konsti{a Stokes number of} $\textrm{St}=\Omega \tauS = 0.1$ particles, representing typical \konsti{maximum} pebble sizes in protoplanetary disks \konsti{\citep[see e.g.,][]{Birnstiel2012}}. 

In \citet{Schreiber2018}, we also performed 2D simulations with $\textrm{St} = 0.01$ 
which also confirmed our collapse criterion, yet at much higher computational cost. For now, high resolution 3D simulations at these small $\stokes$ numbers are not feasible. 

The resolution is $128$ cells per dimension. With on average 10 particles per cell, this leads to a total number of $20,971,520$ particles. For the following collapse simulation, we needed 1.4 $\times 10^{6}$ core-hours on 1024 cores in parallel, a total of 58 days of net running time, spread over one year. Larger numbers of cores would not help for such a small system. 
The 3D parameter study on the SI with various resolutions, initial dust-to-gas ratios and Stokes numbers to identify a suited setup for our simulation \citep{SchreiberThesis} consumed another 15 $\times 10^{6}$ core-hours, without which the simulations presented here could not have been performed. This is only to justify that we did not do an extended parameter study as we did in our two-dimensional study \citep{KlahrSchreiber2020a}.

The pressure gradient was set to $\beta=-0.1$, which translates into sub-Keplerian speed $dV$ as
\begin{equation}
dV = \eta v_K = -\frac{1}{2} \beta c_s.   
\end{equation}
\corrected{This pressure gradient is twice as large as then one used in \citet{JohansenYoudin2007} in order to compensate for the fact that our box is 100 times smaller and does not cover the fastest growing modes of SI, but still driving the SI to a saturated level of diffusion within the given computation time. In the next Section we will directly compare the diffusivities measure in our simulation with the one in \citet{JohansenYoudin2007}.}
The initial dust-to-gas ratio is set to $\varepsilon_0=1$ same as in \citet{JohansenYoudin2007}, and as such chosen three times lower than in \citep{KlahrSchreiber2020a}. Although dust-to-gas ratios greater than $\varepsilon_0=1$ do not fundamentally change the nature of SI, they need longer computation time to reach saturated turbulence \citep{Schreiber2018}.

\begin{figure*}
\plotone{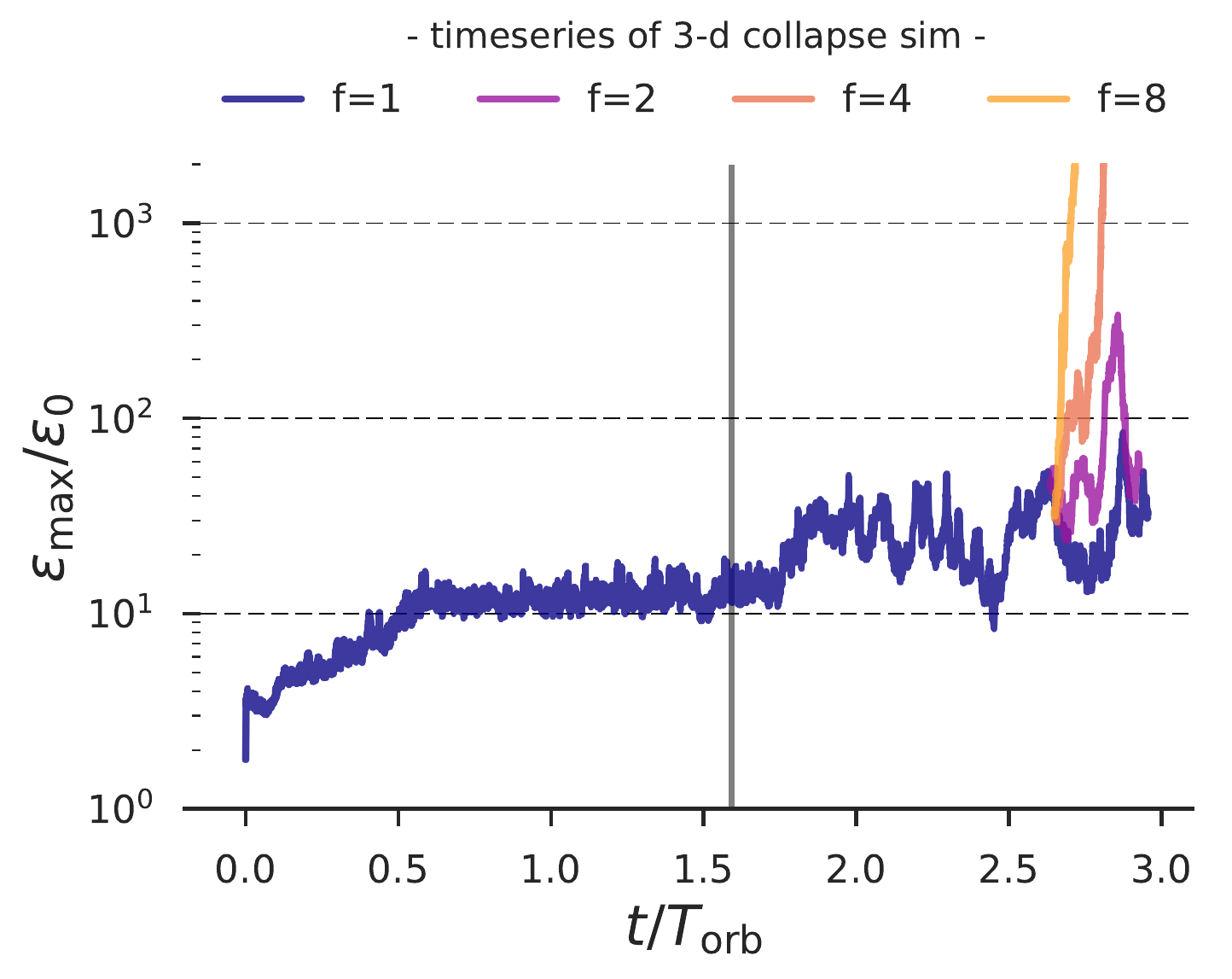}
\caption{Timeseries of the maximum dust-to-gas ratio in the 3D collapse simulation with $St=0.1$ particles in units of the average dust-to-dust ratio $\varepsilon_0$. The grey vertical line indicates the time when self-gravity is turned on.  The initial gravity parameter is set to $\tilde{G}_0=0.71$, which represents Hill density (blue curve). 
Since no collapse occurred in the initial simulation ($f=1$), the gravity parameter was increased to $\tilde{G} = f\tilde{G}_0$ (other colored lines) to step wise double the mass of pebbles and gas in the box, restarting from a gravitoturbulent snapshot. The simulation with eight times larger mass collapsed immediately (yellow) within one contraction time $\tauC$. The one with four times higher gravity took longer, but collapsed after a quarter of an orbit (orange). The run with two times larger mass did not collapse (purple) for more than 30 contraction times $\tauC$. \label{fig:3d_coll_rhopmax_ts}}
\end{figure*}

\subsection{3D Local SI without Self-Gravity}
The initial run \texttt{mod0} without self-gravity took 22 days of effective run time to bring the simulation into a state of saturated SI, even we already had increases the radial pressure gradient by a factor of two. Hence, all further tests (with increased self-gravity) were performed with this single simulation as its basis. The timeseries of the maximum dust-to-gas ratio is shown in Figure \ref{fig:3d_coll_rhopmax_ts} (blue).
The vertical grey bar indicates when self-gravity is switched on. 
The particle diffusivity is measured in the saturated SI state in radial and vertical direction before self-gravity was switched on.

To measure the radial and vertical diffusion in our simulations, we trace individual particles and fit their increasing displacement with the diffusion ansatz $dx = \sqrt{D t} $. For details of this procedure, we refer to Section 3 of \citet{Schreiber2018}.

The measured and scaled dimensionless diffusivities ($\delta = D / (c_s H)$) in the simulation without self-gravity in the radial direction are
\begin{equation}
\delta_x = \left( {1.90 \pm 1.22} \right) \cdot 10^{-6},
\end{equation}
and in the vertical direction
\begin{equation}
\delta_z = \left( {7.25 \pm 2.20} \right) \cdot 10^{-9}.
\end{equation}
\corrected{A similar yet weaker anisotropy was already reported in \citet{JohansenYoudin2007} yet for a simulation in a much larger box $L = 0.2 H$ and using a pressure gradient only half as the one chosen here:}
\begin{equation}
\delta_{x, L = 0.2 H} = \left( {1.6 \pm 0.2} \right) \cdot 10^{-5},
\end{equation}
and in the vertical direction
\begin{equation}
\delta_{z, L = 0.2 H} = \left( {2.7 \pm 0.1} \right) \cdot 10^{-6}.
\end{equation}
\corrected{See also \citet{Schreiber2018} for extended two-dimensional studies on the anisotropy of diffusion. The overall strength of our diffusivity is thus an order of magnitude smaller in the radial direction and by two orders in the vertical direction than found by \citet{JohansenYoudin2007}, but this effect is unavoidable for the box-size that we need to test our stability criterion. We will discuss the role of the box size and resolution for simulations of planetesimal formation in the presence of SI in the discussion section.}

\corrected{Note that our numerical experiments are to test the criterion for collapse, which should hold for any diffusivity value, but that actual critical masses for pebble clouds in the solar nebula use the larger diffusivities as inferred from values in the literature \citep{Schreiber2018,JohansenYoudin2007}. Additional determinations of radial diffusion in SI and in the presence of additional turbulence, especially for mixed particles sizes \citep{Schaffer2018}, are unfortunately not available yet.} 

The fact that the radial diffusivity was found to be more than two orders of magnitude larger than the vertical diffusion means that the corresponding estimated critical length-scales differ by one order of magnitude. 
For Hill density, our diffusivities translate into a radial critical length scale via Equation \ref{eq:collapseCrit}, i.e.:
\begin{equation}
\lcritx =1.4 \times 10^{-3} = 1.4 L,
\end{equation}
which does not fit into our domain, and a vertical scale of
\begin{equation}
\lcritz =8.9 \times 10^{-5} = 0.089 L,
\end{equation}
which does fit into the box. More importantly, the vertical Jeans length $l_{\rm Jeans} = 2 \pi l_{cz} = 0.56 L$ also fits into the box, and thus 
we can expect vertical contraction. But note that this does not mean collapse, as there can be no gravitational collapse in one dimension.

So as both the critical length scale and Jeans length in the radial direction are larger than the radial box extent, and we have no measure for diffusion in the azimuthal direction, the contraction will possibly not go beyond forming a layer of half-width $l_{c,z}$.

\corrected{Even so we argue that azimuthal diffusion will be equally important at the onset of a three-dimensional collapse, we are not able to determine this diffusion, before the collapse happens. Without forming already an azimuthaly contracted sheet, the pebble motion will be dominated by Keplerian shear, and a particle tracking is impossible to our understanding. Whether a subtraction of the Keplerian profile, before tracking the pebbles, would lead to useful results has still to be shown.}

If we would now switch on self-gravity, then the Toomre parameter for Hill density can be determined from the radial diffusivity (Equation\ \ref{eq:QF}) and found to be $Q_p = 1.94$, indicating stability against linear self-gravity modes. This means that even if a radial Jeans length would fit into our box, it would be stabilized by the Coriolis force. But note that self-gravity might modify the strenght of diffusion, which we are going to study in the next section.

\begin{figure*}
\gridline{\fig{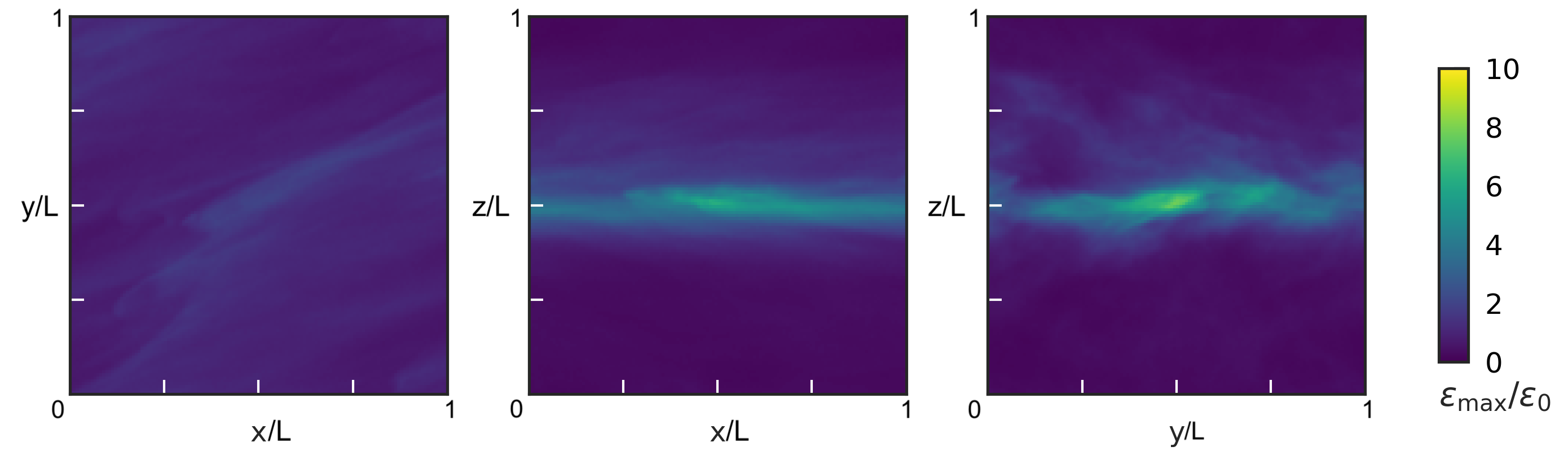}{0.9\textwidth}{(a): $\rho_0 = \rho_{\rm Hill}$ ;  $t = {2.94}{\Omega^{-1}}$}
}
\gridline{\fig{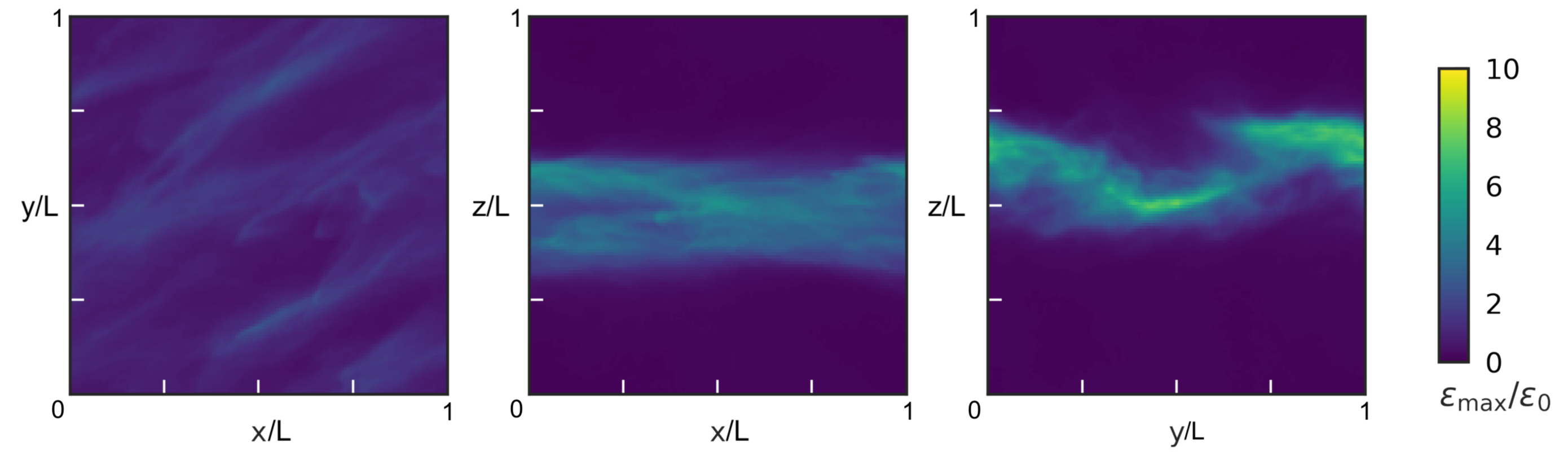}{0.9\textwidth}{(b): $\rho_0 = 2\rho_{\rm Hill}$ ; $t = {2.89}{\Omega^{-1}}$}
}
  \caption{Simulation end-states of the 3D collapse study: Not-collapsing cases. Each column is a projection in a different direction: vertical $=z$ (left), azimuthal $=y$  (middle), and radial $=x$ axis (right). 
  The color bar shows the average dust-to-gas ratio along the projection axis. The simulations have the same set of parameters, only the total mass, and thus the strength of self-gravity is altered via the $f$ parameter.\label{fig:3d_coll_endstatesa}}
\end{figure*}

\subsection{A simulation at Hill density}
At 1.6 orbital periods, after SI has saturated we switch on self-gravity (See Figure\ \ref{fig:3d_coll_rhopmax_ts}).
We set the dimensionless gravity constant in the Pencil Code to $\Gmod \equiv G f \rho_{\rm Hill}/\Omega^2 = 9/4 \pi = 0.71$, which by this construction means the average density in our simulation domain is for $f=1$ exactly the Hill density.

Over the next orbital period we noticed that the density fluctuations increased (see Figure\ \ref{fig:3d_coll_rhopmax_ts}), but no collapse and planetesimal formation happened in this run with $f=1$. This is of particular significance because the self-gravitating clumps in our simulation exceed Hill density by a factor of 30, yet are still not able to contract against the turbulent diffusion. 
\corrected{We continued the simulation for a total of 1.4 orbits, which for the average density (Hill density) corresponds to 14 free fall times $\tauFF$ calculated as:
\begin{equation}
	\label{eq:freefalltime}
	\tauFF = \sqrt{\frac{3 \pi}{32 G \rho}}\, \approx 0.1 \Torb \sqrt{\frac{\rhoHill}{\rho}}.
\end{equation}
If we consider the average over-densities in the simulation of 30 times Hill density (See Figure \ref{fig:3d_coll_rhopmax_ts}), then we even ran the simulation for 75 free fall times.
As we discuss in \citet{KlahrSchreiber2020a} contraction time $\tauC$ for a pebble cloud with $\tauFF \Omega > St$ is actually longer than the free fall time because of the friction of dust with the gas:
\begin{equation}
	\tauC = \tauFF \left(1 + \frac{8 \tauFF}{3 \pi^2 \tauS}\right).
	\label{eq:full}
\end{equation}
In that case the 1.4 orbits would correspond to 5.2 contraction times at Hill density, yet if we again use the average clump density of 30 Hill densities, then we find that we ran our simulation for about 58 contraction times without a gravitational collapse happening.
We are therefor confident that even for longer run times no collapse would have occurred.}

Instead, a vertically contracted particle layers form. Note that these dust layers are not necessarily at the disk midplane, as we do not include vertical stellar gravity in our simulations defining such a midplane. 
The existence of these layers conforms with our prediction in the previous section, where the vertical Jeans length measured from vertical diffusivities was shown to be smaller than our domain size.

Independent of whether such a vertical contraction was due to sedimentation in the stellar gravity field or by the gravitational potential of the pebbles themselves, such a dust layer will modify the SI \citep{Johansen2009} and trigger additional instabilities, such as Kelvin Helmholtz instability (KHI) \citep{JohansenHenningKlahr2006}. While,  \cite{Bai2010b} suggest that the SI might change its behavior and become stronger with sedimentation,
\citet{Gerbig2020} show that the KHI may be as important as the SI, due to setting the vertical extent of the particle layer.
To verify whether or not the KHI is active in our simulation we determine the Richardson Number for our particle layer, which expresses the ratio of stabilizing vertical stratification with the vertical shear. For that, we need to characterise the dust layer.

\corrected{The pebble distribution itself is too noisy to directly measure the local Richardson number for the flow \citep{JohansenHenningKlahr2006}. Therefore, we determine the variance of the pebble layer density to determine its thickness.}
Thus we fit the dust layer with a Gaussian distribution in the spirit of a non-selfgravitating pebble layer, even though we argued that the correct analytic solution for constant diffusion and a vertically infinite domain would be a hyperbolic function (See Equation\ \ref{eq:cosh}), because the scale height of Gaussian and hyperbolic function are quite similar as shown in \citet{KlahrSchreiber2020a} and the standard deviation of the particles is a clearly defined value.

Hence, we measure the scale height via the standard deviation of the vertical dust-to-gas ratio distribution to be $\hlcritz = 1.31 \cdot 10^{-4} H$. So about $7$ scale heights fit vertically into our simulation domain, about $1.5$ times as thick as $l_z$ without sedimentation.
This scale height can directly be translated into a new vertical diffusivity (see Equation\ \ref{eq:firstlc3}) and we get 
\begin{equation}
 \delta_z = St \, \frac{9 }{2 \sqrt{2}} \frac{\hlcritz L}{H^2} = {4.2} \cdot 10^{-8},
\end{equation}
with an increased midplane density to about $\rho_c = 2.7 \rhoHill$. 
We find that the inclusion of self-gravity increases vertical diffusivity by about a factor of 6. With the vertical structure and the midplane pebble to gas ratio we can now determine the Richardson number \citep{Chandrasekhar1961} as function of the combined pebble and gas density $\rho^* = \rho + \rho_g$ as:
\begin{equation}
\mathrm{Ri} \equiv \frac{\left({g_z}/{\rho^*}\right)\left(\partial \rho / \partial z\right)}{\left(\partial v_\phi/\partial z\right)^2},
\label{eq:Ri_general}
\end{equation}
in which for the vertically isothermal ansatz, e.g. vertically constant diffusion, we can replace gravity by stratification $\rho g_z = - D \partial \rho / \partial z$ and $\mathrm{Ri}$ reduces to: 
\begin{equation}
\mathrm{Ri} = \frac{\delta}{St} c_s^2 \frac{\rho}{\rho^*} \frac{\left(\partial \ln \rho / \partial z\right)^2}{\left(\partial v_\phi/\partial z\right)^2}.
\label{eq:Ri_general1}
\end{equation}
The azimuthal velocity from the Nakagawa solution \citep{Nakagawa1986} as a function of the dust-to-gas ratio and radial pressure gradient $\eta$ is $v_\phi = v_{\phi 0} - (\beta/2) (\rho_g/\rho^*) c_s$, so we find 
\begin{equation}
\partial v_\phi/\partial z = \frac{\beta}{2} \frac{\partial \rho^* / \partial z}{\rho^{*}} \frac{\rho_g}{\rho^*} c_s.
\end{equation}
Thus, the Richardson number is:
\begin{equation}
\mathrm{Ri} = \frac{\delta}{St} \frac{4}{\beta^2} \left(1 + \varepsilon\right).
\label{eq:Ri_general2}
\end{equation}
In the case where self-gravity dominates, the pebbles are responsible for each ingredient in the Richardson number and there is no dependence on height in this expression. Also in contrast to the non-self-gravity case \citep{Chiang_2008, Gerbig2020} the value is proportional to $\varepsilon + 1$ and not $\left(\varepsilon + 1\right)^3$, because here the dust-to-gas ratio defines vertical gravity.

For an average midplane density of $\rho_0 = 2.7 \rhoHill$, we thus find $\mathrm{Ri}(f=1) = 0.3$, between the hypothetical critical Richardson number for KHI $\mathrm{Ri}=0.25$ and the numerically determined value of $\mathrm{Ri} \approx 1$ \citep{JohansenHenningKlahr2006}, indicating KHI is likely active here. 
Nevertheless, without more simulations of self-gravity including SI and KHI it will be difficult to distinguish the role of the two instabilities for our chosen scenario.
Yet, for the purpose on how much turbulent diffusivity is needed to prevent planetesimal formation, or more precisely, how much mass is needed to overcome a certain level of turbulent diffusion, this question is irrelevant.

We also found that radial particle diffusion increases within the emerging self-gravitating dust layer. In contrast to the vertical diffusion value, the new radial diffusion value $\delta_x$ can be measured with the default method of tracking the particle travel distance over time. The new value for the radial particle diffusion is then an order of magnitude stronger than without self-gravity, i.e.\
\begin{equation}
\delta_x = \left( {2.38 \pm 1.38} \right) \cdot 10^{-5}.
\end{equation}
This means that the pre-existing anisotropy in diffusion in the absence of self-gravity is preserved. See table \ref{tab:simdata0}. \corrected{The radial diffusivity value is now at the same level as for the larger boxes without self gravity \citep{Johansen2009}, while the vertical diffusion is still about 64 times weaker, meaning a factor of 8 in our length-scale estimates.}

For our setup at Hill-density, i.e. for $f = 1$,
we find no gravitational collapse (as seen in Figure\ \ref{fig:3d_coll_rhopmax_ts}).
Instead, the gas and pebble mixture develops stronger turbulence and the amplitude of 
density fluctuations increases. 
\corrected{In comparison to simulations without self gravity, where the SI formed multiple filaments in the vertical direction (see the right frame in the upper row of Figure\ \ref{fig:3d_prep_simulations}), now a single almost plane pebble layer is created as result of pebble self gravity.}

So far our assumptions about a stability criterion seem to find support in the numerical simulation, i.e.\ that if a box is too small in one direction (here the radial) to not have a critical length fit into it $\lcrit > \onehalf L$, then it cannot collapse, despite being at Hill density. But how can we study the critical box size or critical mass from which on our critical length scales would fit into the domain or respectively unstable wavelength fit in?
\begin{figure*}
\gridline{\fig{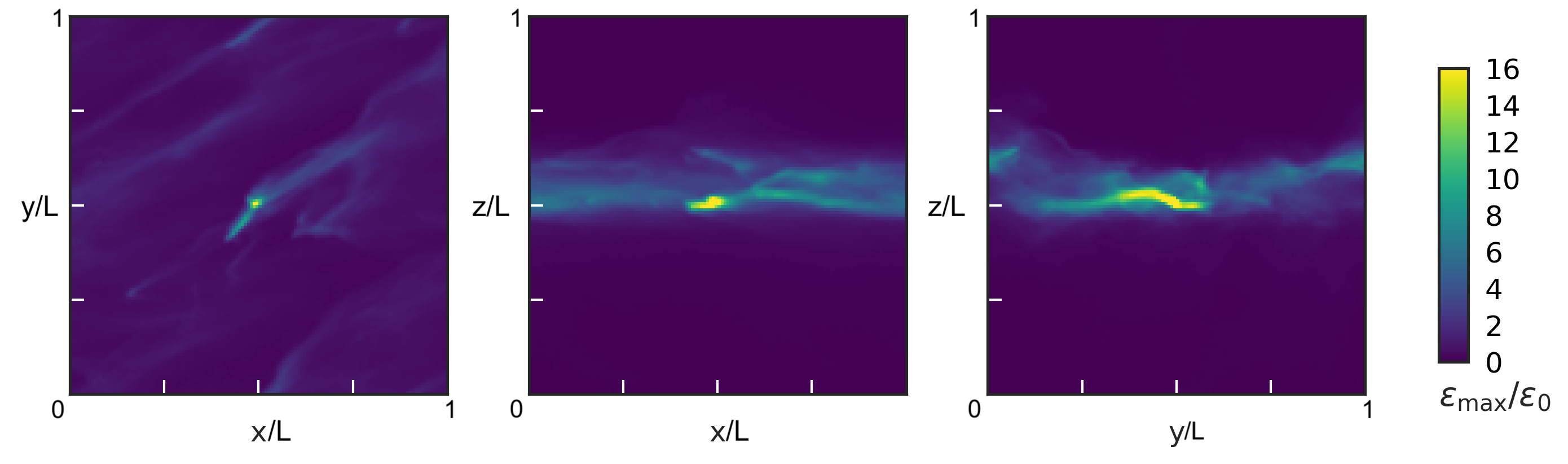}{0.9\textwidth}{(c): $\rho_0 = 4\rho_{\rm Hill}$ ; $t = {2.79}{\Omega^{-1}}$}
}
\gridline{\fig{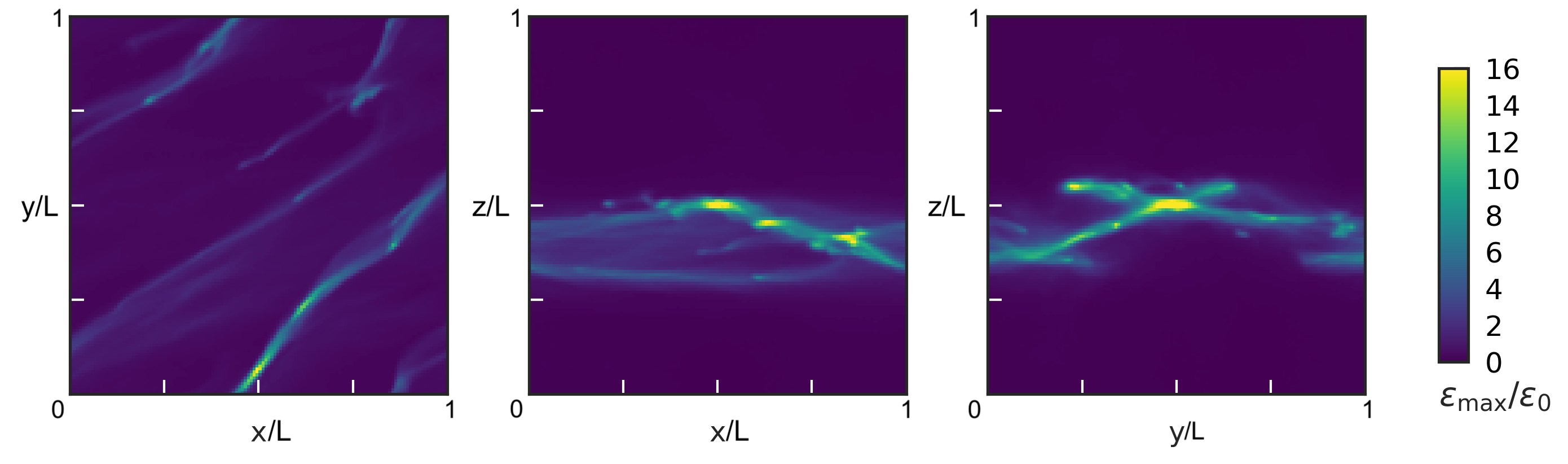}{0.9\textwidth}{(d): $\rho_0 = 8\rho_{\rm Hill}$ ; $t = {2.69}{\Omega^{-1}}$}
}
  \caption{Simulation end-states of the 3D collapse study: Collapse cases. Each column is a projection in a different direction: vertical $=z$ (left), azimuthal $=y$  (middle), and radial $=x$ axis (right). 
  The color bar shows the average dust-to-gas ratio along the projection axis. These two simulations have the same set of parameters as the ones shown in Fig\ \ref{fig:3d_coll_endstatesa}, only the total mass, and thus the strength of self-gravity is altered via the $f$ parameter.\label{fig:3d_coll_endstates}}
\end{figure*}

\subsection{Simulations above Hill density}
Ideally, we would repeat the simulation from scratch with a larger box, however,
this is currently numerically too expensive. We therefore choose to make the critical length-scale smaller by increasing the total mass in the simulations, i.e.\ by decreasing the overall Toomre $\tilde{Q}$. 
Technically we achieve a decreased Toomre $\tilde{Q}$ by increasing the gravitational constant, which is equivalent to up-scaling the total mass in the box. As seen in e.g., \citet{Simon2016, Schaefer2017, Gerbig2020}, who also used this method in their numerical studies of planetesimal formation, this procedure has the advantage of neither directly affecting
the strength of SI, nor, as can be seen in the Richardson number, the strength of KHI, because the average dust-to-gas ratio remains unchanged.

As our numerical simulation is scale free, decreasing $\tilde{Q}$ by a factor of two is similar to doubling the box \corrected{volume (increasing the box dimension in terms of the critical length $L/\lcrit$ by $25\%$) especially as long as we are gravity dominated. 
In case of pure SI, diffusivity would increase with an actually larger box size, and we would have to recalculate the now larger $\lcrit$ \citep{KlahrSchreiber2020a}, which would not be possible for collapsing cases. Thus by keeping $L$ and $\varepsilon_0$ as in the $f=1$ simulation, we argue that diffusivity does not change too much with increasing $f=2,4,8$. We indirectly find support for this assumption as the $\hlcrit(f)$ criterion, based on the the diffusivity in the $f=1$ case succesfully describes the outcome of the simulations with $f=2,4,8$.}

\begin{deluxetable}{ccccccccc}
\tablecaption{Simulation Results for $f=0$ and $f=1$.\label{tab:simdata0}}
\tablehead{
\colhead{model} & \colhead{self gravity}& \colhead{$\delta_x$} & \colhead{$\delta_z$} & \colhead{$Q_p$} & \colhead{$l_{c,x}$} & \colhead{$l_{c,z}$} }
\startdata
\texttt{mod0}  & No &$1.9\times 10^{-6}*$ &  $7.3\times 10^{-9}*$  & $1.9$ & $1.5$& $0.09$ \\
\texttt{mod1}  & Yes &$2.4\times 10^{-5}*$ &  $4.2\times 10^{-8}\#$ & $6.9$ & $5.1$ & $0.2$ 
\enddata
\tablecomments{This table collects the simulation results from our two base line models with initially Hill density $f=1$: (1) the name of the model, (2) gravity switch, (3) and (4) are normalised radial and vertical diffusion (5) $Q$ is the particle Toomre value according to Equation\ \ref{eq:particle_toomre_q}. The following length-scales are given in units of the boxsize $L$: (6) $l_{c,x}$ and (7) $l_{c,z}$ are the radial and vertical critical length. Diffusivities denoted with $
^*$ have been measured using the particle tracking, those denoted with $^\#$ by measuring a scale height.} 
\end{deluxetable}

In model \texttt{mod2} we increased the mass in the box by doubling the gravitational constant ($f=2$), leading to a smaller Toomre parameter and shorter critical length scales (see Tab.\ref{tab:simdata1}). Due to the short duration of the run, we were not able to determine a new diffusivity, so we assumed the same diffusivity as before in the ($f=1$) simulation. While both radial and vertical length scales became smaller, particles did not collapse.

\corrected{The highest pebble concentration was $\varepsilon_\mathrm{max} = 400 \varepsilon_\mathrm{0} = 800 \varepsilon_\mathrm{Hill}$, thus the contraction time would have been $\tauC = 5 \times 10^{-3} \Torb$ and even considering for the average maximum pebble load of about 100 $\varepsilon_\mathrm{Hill}$ we have run the simulation effectively for 34 contraction times without a collapse happening. Thus we deem \texttt{mod2} not collapsing for many contraction / collapse times as stable.}

\corrected{Interestingly, the particle filament from the $f=1$ case is now warped in the azimuthal direction reminding of a KHI shape (as seen in the $y-z$ plot), in a similar way as found in simulations including vertical gravity \citep{Gerbig2020}, an indication of the modified SI and specially the KHI modes, as indicated by the Richardson number.}

Only after increasing the mass to $f=4$ (\texttt{mod4}), the pebble cloud starts to fragment, what also happened in the case of $f=8$ (\texttt{mod8}).
In both cases we started from the same gravoturbulent snapshot based on $f=1$ to save computational effort and to mimic a gradual increase in mass load to allow the system to seek for a new stable state if possible. If one would start with self-gravity at these large dust masses in a laminar disk, collapse could occur before turbulence is triggered.

\corrected{In Figure\ \ref{fig:3d_coll_endstatesa}, we show the end states for the two simulations that included self-gravity (\texttt{mod1},\texttt{mod2}) that did not collapse and in Figure\ \ref{fig:3d_coll_endstates} those that did collapse (\texttt{mod4},\texttt{mod8}). We are plotting the averaged dust to gas ratio in the line of sight. Simulations with $f\leq2$ show no fragments, but $f=1$ shows a single prominent elongated cloud, diagonally located in the $x-y$ plane (as seen from the top view) and $f=2$ two distinct elongated clouds (seen best, when comparing the top view $x-y$ with the side view $y-z$) which do not contract further despite strong self-gravity. The two simulations with higher total mass collapsed each into a single planetesimal, though the run with $f=8$ shows some additional overdensities which are unclear if they also would collapse, if we could continue the simulation.}

\corrected{All these filaments are tilted in the $x-y$ plane in the opposite direction to shear. Whereas \texttt{mod0} (see Figure\ \ref{fig:3d_prep_simulations}) shows the typical trailing wave behaviour in the filaments created by the streaming instability, we now see the effect of the self gravity. Using the Hill density and above for the pebbles implies that perturbations will not get sheared out by the tidal forces from the sun. This enables the formation of trailing and leading filaments likewise, as can also be seen in simulations of
gas disks around young stars, in which gravitational bound structures emerge, i.e.\ planet formation via gravitational fragmentation \citep{Durisen2007}.}

We had to stop the still ongoing collapse simulations, when the density started to diverge, limiting our time-step. For the further evolution \corrected{of the contracting pebble clouds into planetesimals} we refer to simulations like those presented by \citet{Nesvorny2019}, who recently studied the formation of binary planetesimals from collapsing pebble clouds.

In neither the $f=4$ or $f=8$ cases, did the radial critical length $\hlcritx$ fit into the box (see table 2). But in all simulations from $f=1$ to $f=4$ the vertical length-scale $\hlcritz$ easily fit into the box. So neither asking that at least one direction is gravitational stable nor to ask that all directions are gravitational unstable seems to be a good criterion for collapse. 
Likewise, the Toomre parameter that we calculated for each run is not an adequate predictor. As defined above (see Equation\ \ref{eq:QF}) $Q_p$ is independent from vertical diffusion in our simulation setup, as we have a fixed surface density of pebbles. Thus $Q_p$ being set by the stronger radial diffusion only falls below unity for the highest mass case (\texttt{mod8}) $Q_p = 0.86$, in which the fastest growing wavelength (see Equation\ \ref{eq:lambda_fgm}) would be $\lambda_{\rm fgm} \approx 0.01 H = 10 L$, certainly not fitting into our domain.  Neither did the smallest unstable wavelength $\lambda_{\rm min} = 0.0067 H = 6.7 L$ fit into the domain, which indicates how large a box would have to be in order to study the classical gravitational instability in a simulation. \corrected{Those scales are covered and  resolved in simulations of large scale planetesimal formation, yet, as the diffusion was not measured in \citet{Li2019} for the large chosen Stokes Number $\stokes = 2$, we can not determine what actual value their Toomre parameter would have obtained.
We can only speculate that the $\lambda_{\rm fgm}^2$ may define the mass of the largest or most abundant planetesimals formed in large scale simulations if we extrapolate from studies of gas disk fragmentation \citep{Kratter2010}, but this is still left to be shown.}

\begin{deluxetable*}{cccccccccccc}
\centerwidetable
\tabletypesize{\scriptsize}
\tablecaption{Simulation Results including self gravity\label{tab:simdata1}}
\tablehead{
\colhead{model($f$)}  & \colhead{$Q_p$} & \colhead{$\hlcritx$} & \colhead{$\hlcritz$} & \colhead{$\hlcrit(\rho_0)$} & \colhead{$a_{\rm box}[km]$} & \colhead{$a_{c}[km]$}& \colhead{$a_{\rm Jeans}[km]$} &\colhead{$\rho_{\rm c}/\rho_0$} & \colhead{$\hlcrit(\rho_c)$} & \colhead{$\rho_{\rm max}/\rho_0$}&\colhead{Collapse?}}
\startdata
(1)  & (2) & (3) & (4)  & (5)    &(6) & (7)  &(8)&(9)&(10)&(11)  &(12)\\
\hline
\texttt{mod1}  & $6.9$ & $5.1$ & $0.2$  & $1$     & $42$  & $71$  & $165$&$52000$ &$0.004$&$90$   & No\\
\texttt{mod2}  & $3.4$ & $3.6$ & $0.15$  & $0.74$ & $53$  & $63$  & $147$&$6500$ &$0.009$ &$300$    & No\\
\texttt{mod4}  & $1.7$ & $2.6$ & $0.1$  & $0.53$  & $66$  & $56$  & $131$&$800$ &$0.018$&$>1000$  & Yes\\
\texttt{mod8}  & $0.8$ & $1.8$ & $0.08$ & $0.37$  & $84$  & $50$  & $117$&$102$  &$0.036$&$>1000$  &  Yes
\enddata
\tablecomments{This table collects all simulation results including self-gravity for different values of initial average pebble density $f = \rho_0/\rhoHill$ in units of a Hill density: The columns are: 
(1) the name of the model, with the number being equal to $f$, (2) $Q$ is the particle Toomre value according to Equation\ \ref{eq:particle_toomre_q}. The following length-scales are given in units of the box-size $L$: (3) $\hlcritx$ and (4) $\hlcritz$ are the radial and vertical critical length. (5) $\hlcritz(\rho_0)$ is the dimensional average critical length for the mean density in the box. Note that $\hlcrit(\rhoHill) = \lcrit$. (6) $a_{\rm box}[km] $ gives the actual pebble mass in the simulation, given in equivalent compressed diameter ({The conversion from mass $m$ to diameter $a[km] $ needs the definition of a nebula model, particular here a local temperature: $H/R = 0.03$, the mass of the central object: $M_\star = M_\odot$, and a solid density for a collapsed body $\rhoSolid = 1 \mathrm{g}/\mathrm{cm}^3$}, (7) $a_c$ is the equivalent diameter for pebble clouds of mass $m_c$, i.e.\ for mean density $\rho_0$, (8) $a_{\rm Jeans}[km]$ is the equivalent diameter for the mass of a contracted B.E.\ sphere (of Jeans Mass and for density at the surface $\rho_\circ = f \rhoHill$), (9) $\rho_{\rm c}/\rho_0$ is the central density peak with respect to the initial density needed to make a B.E.\ sphere of mass $a_{\rm box}[km]$ unstable, which results in (10) a critical length for the sphere of $\hlcrit$ for given central density. (11) gives the maximum density fluctuation achieved in the individual simulations $\rho_{\rm max}$ compared to the initial density,
and the last column (12) states whether collapse occurred. Collapse happens when the critical length fits about twice into the domain ($\hlcrit \lessapprox \onehalf L$).
The central compression needed for collapse is then less than 1000 and can easily be resolved on the grid.
The size predicted on just diffusion and Hill density is $a_c = 71$ km, and falls just squarely between the two collapsing models. Note that other nebula parameters, foremost $\varepsilon_\mathrm{Hill}$ and $H/R$, will lead to other equivalent sizes.} 
\end{deluxetable*}

For the case $f=4$ (\texttt{mod4}), the Toomre value exceeds unity $Q_p = 1.7$ and still the pebble cloud fragments. So as a general outcome, linear Toomre modes are not necessarily indicative of gravitational collapse for our pebble clouds.

So we probably have to consider the three-dimensional shape of the pebble cloud as it results from an-isotropic diffusion and then see what determines the stability of this body. 

In the case of isotropic pressure, or in this case isotropic diffusion we would expect a spherical structure to evolve, in which gravity and diffusion may counteract. In the case of non-isotropic diffusion one would then expect an ellipsoidal structure. This idea has been championed in the context of elliptical galaxies, with anisotropic r.m.s.\ velocities \citep{1981gask.book.....M} and even goes back to \citet{Schwarzschild1908}. As in our simulations diffusion and gravity define one length per dimension, which are our $\hlcritx$, $\hlcritz$ and a so far unknown $\hlcrity$, we can construct an ellipsoid of the follwing shape:
\begin{equation}
1 = \frac{x^2}{\hlcritx^2} + \frac{y^2}{\hlcrity^2}+ \frac{z^2}{\hlcritz^2},
\end{equation}
with the volume 
\begin{equation}
V = \frac{4 \pi}{3} \hlcritx \hlcrity \hlcritz = \frac{4 \pi}{3} \hlcrit^3.
\end{equation}
Thus an elliptic pebble cloud has the same mass as a sphere of radius $\hlcrit$,
which would be result of the individual diffusivities combined in the following way:
\begin{equation}
\delta = \sqrt[3]{\delta_x \delta_y \delta_z} \approx \sqrt{\delta_x \delta_z},
\end{equation}
with the assumption $\delta_y \approx \sqrt{\delta_x \delta_z}$ for lack of ways to obtain this value otherwise.
For the diffusivities as measured for $f=1$ this results in a value of $\delta = 10^{-6}$.

In that case, the resulting critical length $\hlcrit$ roughly fits into our box for the collapsing cases $f=4$ and $f=8$ cases but not in the stable cases $f=2$ and $f=1$ (see Table \ref{tab:simdata1}). 

Furthermore if we calculate the mass of this ellipsoid $m_c$ (see Equation\ \ref{eq:mass}) and express the result as equivalent radius $a_c$ for the individual simulations (see table \ref{tab:simdata1}), then we notice that those cases collapsed in which the total amount of pebbles, expressed in equivalent size $a_\textrm{box}$ is clearly larger than the critical size $a_c$. \corrected{This is a direct confirmation of our collapse criterion $m > m_c$ as defined in Equation\ \ref{eq:mass_simple}.}

\corrected{It may seem adhoc that we base our collapse criterion on the average density of the simulations and on the box size, when on the other hand clearly much smaller and much denser structures do form in the simulations. This can be justified by checking how the radius $\lcrit$ of an unstable pebble cloud (see Equation\ \ref{eq:mass_simple}) scales with its density in units of Hill density, i.e.\ $f$:}
\begin{equation}
    \lcrit(f) \sim f^{-\frac{1}{2}},
\end{equation}
\corrected{thus a pebble concentration 10 times smaller than $\lcrit$ has to be on average at least more than 100 times denser ($f = 100$) than the average density to full-fill the collapse criterion. More importantly the  critical mass in a clump scales as}
\begin{equation}
    m_c(f) \sim \lcrit(f)^3 f = f^{-\frac{1}{2}},
    \label{eq:mcf}
\end{equation}
\corrected{and thus the resulting compressed diameter scales as}
\begin{equation}
    a_c(f) \sim  f^{-\frac{1}{6}}.
    \label{eq:acf}
\end{equation}
\corrected{In other words, unstable fragments smaller than the box size at several times the Hill density, do not represent a significantly smaller compressed size, than the box at Hill density itself.}

\corrected{Following our numerical simulations is seems difficult to create that massive and compact clumps on small scales (more than $10\%$ of the pebbles in $0.1\%$ of the volume), if the large scales are not already unstable. Here we have not yet considered that the actual pebble concentrations are not of constant density, but have an internal stratification, which we will investigate in the next step.}

\section{A Bonnor-Ebert solution for pebble clouds}\label{subsec:3-4}
The original derivation of the $l_c < \onehalf L$ criterion stemmed from a time scale argument, that contraction is faster than diffusion. The assumption was a sphere of constant density, with sharp cut-off boundaries. 

A different approach to derive a critical mass would be study the local equilibrium between gravity and diffusion in a similar fashion as we did for
the midplane layer for the particles. But since collapse needs more than one dimension, we are also motivated to ask what the three dimensional shape and profile of a self-gravitating pebble cloud would be under the influence of internal turbulent diffusion.
As above we replace our ellipsoid with a spheroid of equivalent mass and same central density, yet with the spatial averaged diffusivity.
This can be studied using Equation \ref{eq:plane} but now in spherical coordinates, assuming spherical symmetry, where $r$ is the distance from the clump center, which leads to the Lane-Emden equation: 
\begin{equation}
\rho =  - \frac{D}{\tau} \frac{1}{4 \pi G } \frac{1}{r^2}\partial_r r^2 \partial_r\ln{\rho}.
\label{eq:sphere}
\end{equation}
One can rewrite this in terms of the critical length $\hlcrit$, which is equivalent
to the normalisation value for the dimensionless radius of a Bonner-Ebert sphere (see Equation\ 9.6 in \citet{StahlerPalla2008}) 
\begin{equation}
\frac{\rho}{\rho_c} =  - \frac{\hlcrit^2}{r^2} \partial_r r^2 \partial^2_r\ln{\rho}.
\label{eq:BE}
\end{equation}
The resulting radial profile is the Bonnor-Ebert (BE) solution, which can be found by solving the Lane-Emden equation numerically for an isothermal equation of state with the characteristic radius $\hlcrit$ following \citet{StahlerPalla2008}.
Obviously this BE sphere has the same characteristic scale $\hlcrit$ as the plane layer, which is the same critical length of $\lcrit$ from the time scale criterion, for $\rho_c=\rhoHill$. 

For a given temperature, or in our case for a given diffusivity and Stokes-number combination, a family of different solutions is possible, only depending on the ratio of the sphere's central density $\rho_c$ to the Hill density. If the radially decreasing density falls below $\rho_\circ = \rho_c/14.1$, then the cloud will be linear unstable for collapse. This maximal density ratio of $\rho_c/\rho_\circ = 14.1$ has to be determined numerically and is a general property of the isothermal BE solutions.
The mass of a BE sphere 
is a function of the numerically obtained non-dimensional cloud mass $m$ as a function of density contrast and the pressure at the surface $P_\circ$
\begin{equation}
M_{\rm BE} = \frac{m a^4}{P_\circ^{1/2} G^{3/2}} = m \frac{\left(\delta/St\right)^{3/2} c_s^3}{\rho_\circ^{1/2} G^{3/2}} 
\label{eq:BE_Mass}
\end{equation}
where we use that our pressure is $P_\circ = a^2 \rho_\circ$ with the equivalent speed of sound $a = \left(\delta/St\right)^{1/2} c_s$ (see Equation\ \ref{eq:asound}).
The critical dimensionless cloud mass was numerically determined as $m_1=1.18$ 
which then defines the critical BE mass also known as the
Jeans mass \citet{StahlerPalla2008}. For reaching the Hill density at the surface $\rho_\circ = \rhoHill$ this leads to
\begin{equation}
M_{\rm Jeans} = m_1 \sqrt{\frac{4 \pi}{9}} \left(\frac{\delta}{St}\right)^{3/2} \left(\frac{H}{R}\right)^3 \left(\frac{\rhoHill}{\rho_\circ}\right)^{1/2}  M_\sun, 
\label{eq:BE_Mass3}
\end{equation}
which already shows the same functional dependence on $H/R$, $\delta / St$ as Equation\ \ref{eq:mass} for $m_c$, which was derived for a sphere of constant density at the Hill value.
\begin{equation}
M_{\rm Jeans} = 12.5 \left(\frac{\rhoHill}{\rho_\circ}\right)^{1/2} m_c, 
\label{eq:BE_MassX}
\end{equation}
This means the equivalent compressed size $a_c$ would be 2.3 times larger for a BE \corrected{solution with a central density of $\rho_c = 14.1 \rhoHill$  when compared to a sphere of constant Hill density}.

\corrected{The radius of the BE solution beyond which it is unstable is $l_\circ = 6.5 \hlcrit$ \citep{StahlerPalla2008}. At this distance the local density drops to $\rho(l_\circ) = \rho_\circ$. Thus the minimum extent of the BE solution to become unstable would be $l \ge l_\circ$. But then the BE sphere (see Equation\ \ref{eq:BE_MassX}) would have a size of $l_\circ = 1.7 \lcrit$ and thus not fit into boxes of $L = 2 \lcrit$.}

\corrected{With increasing central density, both size and mass of the BE solution do decrease. A BE sphere that fits perfectly inside our constant density cloud ($l_\circ = \lcrit$) would then need a central density of $\rho_c = 42 \rhoHill$ and as a result have about $7.2 m_c$ in mass, or correspondingly still about twice as large as described by $a_c$.  Therefore, the unstable BE solutions in our numerical experiments with a limited mass reservoir will need even higher central densities.} 

If we compare the Jeans masses (expressed as equivalent size $a_\mathrm{Jeans}$) for our 4 simulations $f=1$ to $f=4$ (see Table \ref{tab:simdata1}) with the mass in the box $m_\mathrm{box} = f \rhoHill L^3$ and $a_\mathrm{box} = \left({m_\mathrm{box}}/{\rhoSolid}\right)^{1/3}$, we see that it is always larger than the total mass of available pebbles on our simulations domain ($a_{\rm Jeans} > a_{\rm Box}$). But, if we express the Jeans mass as a function of its central density, then we can calculate a critical density at which a new Jeans mass $\hat{M}_{\rm Jeans}$ equals the mass of the pebbles in the box. Replacing the density at the surface $\rho_\circ$ with the central density $\rho_0 = 14.1 \rho_\circ$ we find:
\begin{equation}
\hat{M}_{\rm Jeans} = 47 \left(\frac{\rhoHill}{\rho_c}\right)^{1/2} m_c.
\label{eq:BE_Mass33}
\end{equation}
If we now set $M_{\rm Box} = \hat{M}_{\rm Jeans}$ we can calculate the necessary density to make this cloud unstable 
\begin{equation}
\rho_c = \left(\frac{47 m_c}{M_{\rm Box}}\right)^2 \rhoHill
\label{eq:BE_Mass4}
\end{equation}
and compare this value to the average density in the box $\rho_0$.
For the cases $f=1$ and $f=2$, overdensities in $\rho_c / \rho_0$ of a factor of more than 10000 are needed to make a BE sphere of given mass $M_{\rm Box}$ collapse. The necessary resolution for the central peak at size $\hlcrit(\rho_c)$ would hardly be reached. Yet for models $f=4$ and $f=8$ a concentration of only about 1000 of the initial density is needed to trigger collapse, and these are the peaks in pebble concentration that we observe in the collapsing runs see Figure\ \ref{fig:3d_coll_rhopmax_ts} when converting the plotted $\varepsilon / \varepsilon_0$ into $\varepsilon / \varepsilon_\textrm{Hill}$
\begin{equation}
\varepsilon = f \varepsilon_0.
\label{eq:BE_Mass5}
\end{equation}
Based on Equation\ \ref{eq:BE_Mass33} we can now see that one needs to increase the 
central density fluctuation by $10^6$ to decrease the Jeans mass by a factor of a 1000 and thus producing an equivalent size $a_c$ to be 10 times smaller. This is what makes it so hard to form small planetesimals.

\begin{figure*}
    \centering
    \includegraphics{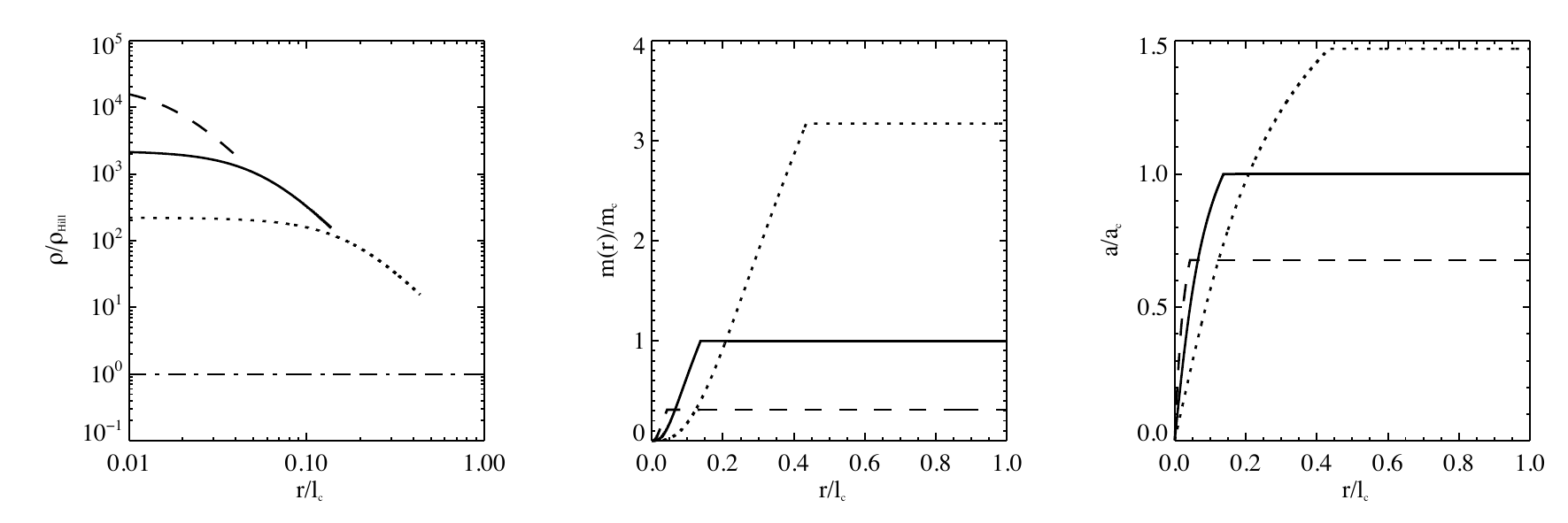}
    \caption{Bonnor-Ebert solutions for pebble clouds. The solid lines depict a solution that has the same mass $m_c$ as a constant density pebble cloud at $\rhoHill$ of radius $\lcrit$ (depicted as dash-dotted line in the left figure). The x-axis is always given in units of that critical length. Left plot: density profile. The pebble density decreases to the critical value $\rho_\circ = 14.1 \rho_c/$ within $0.14 \lcrit$. For 10 times larger (smaller) central density, the profile is getting steeper (shallower) as plotted in dashed (dotted) curves. In the middle plot we show the enclosed mass of the BE solutions. The fiducial case (solid line) matches the desired mass of the constant density solution $m_c$, but the lower density case (dotted line) exceeds the mass by a factor of three and the increased density case (dashed line) would allow for three times smaller masses to collapse. In the right plot we add the equivalent sizes for the three different central density cases and find all to lie within a narrow range around the nominal size $a_c$.}
    \label{fig:4}
\end{figure*}

\corrected{We can illustrate this effect in Figure \ref{fig:4}. We first calculate a BE solution for $\hat{M}_{\rm Jeans} = m_c$, i.e.\ the mass as predicted by the time scale argument. Then it follows that $\hlcrit = \lcrit / 47$ and a central density of $\rho_c = 47^2 \rhoHill = 2209 \rhoHill$. This BE sphere has then a size of $l_0 = 6.5 \hlcrit = 0.14 \lcrit$ and contains the same mass as a sphere of constant density $\rhoHill$ and radius $\lcrit$. In other words, if we reshape our pebble cloud to create a less than a $2 \times 10
^3$ central density increase then it will still be stabilized by diffusion. We also add a profile (dotted line) where the central density peak is 10 times lower than the critical value. This BE sphere would then also fit into our original constant density cloud, but now the BE sphere would be about three times more massive. 
A weaker density bump can only be unstable for larger and more massive pebble clumps. We can also show how strong the density peak would have to be ($\rho_c =  2.2 \times 10^4 \rhoHill$), in order to make a three times lower mass cloud unstable (dashed line).
Finally we also convert those cloud masses into equivalent radii and find that by varying the central density by a factor of 10, only changes the equivalent diameter by a factor of $1.4$ with respect to the prediction based on a constant density sphere.}

\corrected{This strong correlation between the available pebble mass and the necessary over-density for collapse is the reason that our simple time-scale based argument for the critical mass of a constant density sphere at Hill level holds even for numerical simulations with density fluctuations of several thousand and associated length-scales of less than $10\%$ of the simulation domain.}

In Table 2 we summarise the results of our simulations. 
\corrected{All models cover the same volume and the same solid to gas ratio. Therefore also the strength of SI will be similar in these simulations, even we are not able to determine actual diffusivities from the models \texttt{mod2}-\texttt{mod8}. Therefor the assumed diffusivity is still the one from \texttt{mod2}. Based on that diffusivities (see \ref{tab:simdata0}) we determine the pebble Toomre value and the critical length-scales $\hlcrit,\hlcritx,\hlcritz$ for the increased average pebble density. We translate the pebble mass in the simulation to an equivalent compressed mass by assuming a central object of solar mass, a local temperature equivalent to $H/R = 0.03$ and a compressed density for planetesimals of $\rhoSolid \approx 1 \mathrm{g}/\mathrm{cm}^3$.}
\corrected{We also derive the predicted critical equivalent size $a_c$ of the pebble cloud for the measured diffusivities as well as the associated equivalent size for a Bonnor-Ebert solution with Hill density at its outer edge. We find that those models produce a collapse in which the critical length fits into the simulation box $\hlcrit \lessapprox \onehalf L$ (\texttt{mod4,mod8}). In those cases also the density spike of an associated BE sphere made of all available pebbles in the box needs an amplitude of less than $1000$ with respect to the initial pebble density. Such a compression to a scale corresponding to $10\%$ of the box size is easily achieved in the simulations and also still well resolved.} 

\section{Summary, Conclusion and Outlook}
\label{sec:4}
\corrected{In this paper we tested our criterion for the gravitational collapse of a pebble cloud with internal turbulent diffusion. This criterion, as derived in \citep{KlahrSchreiber2020a}, defines a minimum mass $m_c$ for which the contraction time would be faster than turbulent diffusion. We find that for a given value of diffusion pebble clouds exceeding a given mass can collapse, whereas as lower mass clouds will be dispersed.}

\corrected{This collapse criterion is different from asking for the necessary pebble load ("metalicity" or local pebble-to-gas surface density) in order to trigger streaming instability and planetesimal formation in the first place, as we did in \citet{Gerbig2020}. In that paper we were approaching planetesimal formation from the large scales, asking for sedimentation in the presence of turbulent diffusion sufficient to allow for the formation of regions exceeding the Hill density. 
Whereas in \citep{KlahrSchreiber2020a} and the present paper we ask whether all regions of Hill density will automatically collapse into planetesimals or whether there is a mass threshold, like a Jeans criterion.}

\corrected{In the present paper we tested the collapse criterion}
\begin{equation}
m > m_{c} = \frac{4 \pi}{3} \lcrit^3 \rho_{\rm Hill} = \frac{1}{9} \left(\frac{\delta}{\St}\right)^{\frac{3}{2}} \left(\frac{H}{R}\right)^3 M_\sun,
\label{eq:massrep}
\end{equation}
\corrected{in three-dimensional simulations to successfully describe which models would be too low in mass to produce a collapse.}

\corrected{The predicted equivalent sizes $a_c$ of an unstable pebble cloud in the solar nebula or any protoplanetary disk, are set by the strength of the streaming instability, which in turn depends mostly on the local pebble-to-gas ratio at reaching Hill density. The gas mass of a proto-planetary disk is therefor responsible for the resulting planetesimal sizes. 
In the present paper we have chosen a parameter set in particle size, average pebble to gas ratio, gas mass, and pressure gradient, suited to test our collapse criterion $m > m_c$ and we find diameters of $a_c = 70$ km. Other parameters in terms of local pressure scale height $H/R$, pebble sizes, local gas density and pressure gradient will lead to a wide range of sizes. Nevertheless, for the solar nebula model as derived in \citet{Lenz2020} we find equivalent diameters for the pebble clouds of $60 - 120$ km at early times, and values as low as $6 - 12$ km as the nebula disperses.} 

Our simulations focused on the interaction of streaming instability, Kelvin-Helmholtz instability and self-gravity on the scales of planetesimal formation. Therefore we simulated only a small Section of the disk and ignored vertical gravity and effects of large scale turbulence on small scales (as argued for in \citep{KlahrSchreiber2020a}).

\corrected{In agreement with simulations using larger boxes \citet{JohansenYoudin2007} and \citet{Schreiber2018} we find that diffusion by SI is anisotropic. Radial diffusion at small scales is about an order of magnitude stronger than the vertical diffusion, already without self-gravity.
But for the first time we could show how diffusivity (especially in the radial direction) changes after the inclusion of self-gravity.}
The turbulence and diffusion in the self-gravity case is still anisotropic and in both directions significantly larger than in the case without self-gravity. We attribute this increase to a modified streaming instability and to \corrected{potentially} active Kelvin-Helmholtz instability because of vertical self-sedimentation. We measure a Richardson number of $\mathrm{Ri} = 0.3$, well in the possible regime for KHI.

\corrected{Our results provide an explanation for the turn over in the planetesimal size distribution towards small objects ($a_c < 100 \mathrm{km}$) as found in global simulations \citep{Johansen2009, Simon2016, Simon2017, Abod2018}, for which unfortunately the diffusion was not determined. Future simulations will have to clarify the range of diffusivities that can be expected for realistic pebble sizes and pebble size distributions \citep{Schaffer2018}.}

\corrected{The highest resolution studies of SI and planetesimal formation to our knowledge are those by \citet{Li2019} who found at a resolution of $\delta x = 1.9 \times 10
^{-4} H$ a turnover of the size distribution at 100 km diameter for $\stokes = 2$ particles. As diffusivity (especially in the radial direction) was not measured in these runs, a determination of the critical length scales and our pebble cloud collapse criterion, is not possible. Yet, if we use diffusion measurements from \citet{JohansenYoudin2007} and assume that larger particles produce stronger diffusion and thus the ratio of $\delta/\stokes$ may remain relatively constant, we only would have to look at the dust to gas ratio at Hill density in their simulation for the chosen constant of gravity in code units $\Gmod = 0.05$ to make a guess on expected critical length scales. Following \citet{Li2019} the Hill density in code units in their simulation is determined by the constant of gravity in code units as:}
\begin{equation}
\varepsilon_\mathrm{Hill} = \frac{9}{\Gmod} = 180,    
\end{equation}
\corrected{for which our predicted equivalent diameter would be $a_c = 25$ km. Yet it has been reported that the inclusion of vertical gravity can increase diffusivity, thus if the turn-over corresponds to $a_c = 100$ km, this would require a radial diffusion of 16 times stronger than in the unstratified case of \citet{JohansenYoudin2007}. To clearly relate the turnover size to the diffusion limited size and thus the strength of turbulent diffusion, the measurement of the diffusivities in such large scale simulations seems to be unavoidable.}

\corrected{We derived a novel Toomre value for the self gravitating pebble sub-disk under turbulent diffusion. We could show that diffusion leads to a pebble sound speed $a$ that is not the r.m.s.\ speed $v_\mathrm{rms}$ of pebbles, but represents the "pressure" like resistance of pebbles against compression, driven by diffusion:}  
\begin{equation}
 P / \rho = a^2 = \frac{D}{\tauS c_s^2 + D} c_s^2.
\end{equation}
\corrected{Only in the special case that the correlation time $\tau_\mathrm{t}$ of the sufficiently subsonic turbulence is equal to the stopping time $\tauS$ of the particles one would find $a \approx v_\mathrm{rms}$ as with $D \approx v_\mathrm{rms}^2 \tau_\mathrm{t}$ \citep{YoudinLithwick2007} we get}
\begin{equation}
 a^2 = v_\mathrm{rms}^2 \frac{\tau_\mathrm{t}}{\tauS} \frac{1}{1 + \frac{v_\mathrm{rms}^2}{c_s^2} \frac{\tau_\mathrm{t}}{\tauS}} .
\end{equation}
\corrected{All our analyses were based on the assumption that the gas is quasi incompressible. In fact with our box size of $L = 0.001 H$ the sound crossing time is the shortest time scale in the system and was therefor making or simulations so numerical expensive. An incompressible code would have performed much better, but we had none available that also had self gravitating and friction coupled particles incorporated. We find the gas density to fluctuate on a $1\%$ level, which is close enough to incompressibility.}

\corrected{The derived Toomre parameter $Q_p$ for the pebbles is related to the Toomre parameter of the gas disk $Q_g$ as 
\begin{equation}
Q_p = \sqrt{\frac{\delta_x}{\stokes}} \frac{Q_g}{Z},
\end{equation}
which is almost identical to the stability parameter as derived in \citet{Gerbig2020}: \begin{equation}
\tilde Q_p = \frac{3}{2} \sqrt{\frac{\delta_z}{\stokes}} \frac{Q_g}{Z}
\end{equation} 
The difference lies in using radial diffusion $\delta_x$, to study the onset of a linear gravity mode from constant back ground density or using the vertically measured diffusion $\delta_z$ for a collapse criterion in a non-linear state, based on the original $m_c$ criterion. Only for isotropic diffusion $\tilde Q_p$ could be a good approximation for the Toomre value, but in any case $\tilde Q_p\leq 1$ is the condition to get planetesimal formation started by setting a minimal pebble enhancement $Z$ for a given macroscopic diffusion $\delta_z$ (or even $\alpha$) to reach Hill density in the midplane.}  

We have shown that the radial diffusion decides on the Toomre value, whereas the 
vertical diffusion regulates the midplane density. Therefor it is possible to exceed Hill density in the midplane (for vertical diffusion weaker than radial diffusion) without being Toomre unstable.

Gravito turbulence (driven by linear self-gravity modes of the pebbles) may not play a major role since the Toomre value $Q_p$ for the pebble layer is always very large for the measured diffusion strength and even in the one possibly unstable case $Q_p < 1$ no unstable modes fit into the box the linear instability will be suppressed and outgrown by the non-linear collapse.
One only should expect significant gravitational turbulence below $Q_p<1.5$ and according to table \ref{tab:simdata1}, especially the not-collapsing simulations have larger Toomre values.

Therefore it is probably self-sedimentation and the hydro dynamical instabilities that create the gravitational finite amplitude unstable overdensities. 

\corrected{Whereas \citet{KlahrSchreiber2020a} was testing this criterion in vertically integrated two-dimensional simulations of the SI we were performing three dimensional simulations in the present paper. Both two-dimensional and three-dimensional simulations confirm the collapse criterion, if one defines a dimensionally averaged diffusivity $\delta = \sqrt{\delta_x \delta_z}$. In the first paper we performed simulations for different Stokes numbers, different box sizes $L$, we varied the pressure gradient and also the initial dust to gas ratio. In the present paper we kept all those parameters fixed, but we changed the total pebble mass and therefor the critical length $\hlcrit$ even if $\delta$ is not changed.
As a result we were able to perform simulations for $\hlcrit < L/2$, which collapsed, as well as for $\hlcrit > L/2$, which did not collapse for many contraction times, confirming our stability criterion.}

We also derived a Bonnor-Ebert model for pebble clouds 
in equilibrium between diffusion and contraction. In that case, one expresses the mass of the pebble cloud in terms of central density $\rho_c$ (or outer density $\rho_\circ$) and diffusion per Stokes Number.
The equivalent size depends only weakly on central density $a_c \propto \rho_c^{-1/6}$ explaining why smaller planetesimals are less likely to form, as they need a much stronger density fluctuation, before gravity can take over.

In the two simulations in which planetesimals formed (\texttt{mod4} and \texttt{mod8}) we found the onset of a $66$ km and $84$ km pebble heap collapse. These sizes are not unrealistic for planetesimals \citep{Morbidelli2009}, but the collapse was not complete in our simulation, as even at pebble densities of $10^3 \rhoHill$ one is still by a factor of $10^2$ - $10^9$ below the solid density of a planetesimal depending on the location in the nebula. During the further contraction fragmentation into several planetesimals can occur, as a result of the angular momentum of the pebble cloud, as well as erosion by the headwind. Only simulating the further collapse can show how many planetesimals with what size spectrum will form from the collapse of the unstable pebble clouds \citep{Nesvorny2019}. We refer to \citet{KlahrSchreiber2020a} for further discussions on realistic pebble sizes, dust to gas ratios and resulting planetesimal sizes for models of the solar nebula \citep{Lenz2020}, finding a preferred $a_c \approx 100$ km from the asteroid to the Kuiper belt, as argued for by observations \citep{Bottke2005,Nesvorny2011}. 

\corrected{In \citet{Schreiber2018} we find that for high mass loads the strength of diffusion scales inversely with the dust to gas ratio and also proportional to the stokes number, at least over a certain range of pebble sizes. This implies a major dependence of the critical masses on the pebble to gas ratio at Hill density, which can vary strongly over the course of planetesimal formation and a lesser dependence on the pebble stokes number. If SI and diffusion $\delta$ decrease with $\stokes$ then the ratio of $\delta/\stokes$ may stay constant. This effect needs further investigation, especially if one considers a range of stokes numbers as in \citet{Schaffer2018}. The ultimate goal would be to define a representative $a_c$ for a particle size mixture, gas pressure gradients and the local dust to gas ratio at Hill density. And in a second step to learn how the mass spectrum $dN/dM_{\rm P} \propto M_{\rm P}^{-p}$ (see \citep{Johansen2015,Simon2017,Abod2018,Li2019}) of forming planetesimals will relate to this representative $a_c$ and the local availability of pebbles. The results can then be fed into models of planetary embryo and planet formation \citep{Mordasini2009,Johansen2019,Emsenhuber2020,Schlecker2020,Voelkel2020a,Voelkel2020b,Voelkel2020c} and by studying the full model including pebble accretion \citep{KB2006,OrmelKlahr2010,LambrechtsJohansen2012,Lambrechts2019,Bitsch2019,Bitsch2019b,Voelkel2020c} we can ultimately test our paradigm for planetesimal formation in its capability to create the diversity of exoplanets and explain peculiarities of the solar system.}

\acknowledgments{We are indebted to Hans Baehr, Konstantin Gerbig, Christian Lenz and Ruth Murray-Clay for many fruitful discussions and technical advise. Many thanks to our anonymous referee as well as Sanemichi Takahashi and Ryosuke Tominaga to ask the right questions that led us to better understand the concepts of diffusion pressure and its implications for current and future work. This research has been supported by the Studienstiftung des deutschen Volkes, the Deutsche Forschungsgemeinschaft Schwerpunktprogramm (DFG SPP) 1385 "The first ten million years of the Solar System" under contract KL 1469/4-(1-3) "Gravoturbulente Planetesimal Entstehung im fr\"uhen Sonnensystem" and by (DFG SPP) 1833 "Building a Habitable Earth" under contract KL 1469/13-1 \& KL 1469/13-2 "Der Ursprung des Baumaterials der Erde: Woher stammen die Planetesimale und die Pebbles? Numerische Modellierung der Akkretionsphase der Erde." This research was supported by the Munich Institute for Astro- and Particle Physics (MIAPP) of the DFG cluster of excellence "Origin and Structure of the Universe and in part  at KITP Santa Barbara by the National Science Foundation under Grant No. NSF PHY11-25915. 
The authors gratefully acknowledge the Gauss Centre for Supercomputing (GCS) for providing computing time for a GCS Large-Scale Project (additional time through the John von Neumann Institute for Computing (NIC)) on the GCS share of the supercomputer JUQUEEN \citep{Stephan:202326} at J\"ulich Supercomputing Centre (JSC). GCS is the alliance of the three national supercomputing centres HLRS (Universit\"at Stuttgart), JSC (Forschungszentrum J\"ulich), and LRZ (Bayerische Akademie der Wissenschaften), funded by the German Federal Ministry of Education and Research (BMBF) and the German State Ministries for Research of Baden-W\"urttemberg (MWK), Bayern (StMWFK) and Nordrhein-Westfalen (MIWF). Additional simulations were performed on the THEO and ISAAC clusters of the MPIA and the COBRA, HYDRA and DRACO clusters of the Max-Planck-Society, both hosted at the Max-Planck Computing and Data Facility in Garching (Germany). H.K. also acknowledges additional support from the DFG via the Heidelberg Cluster of Excellence STRUCTURES in the framework of Germany's Excellence Strategy (grant EXC-2181/1 - 390900948).}




\newpage
\appendix


\section{Scale height of the pebble layer under self gravity}
\label{sec:A}
The vertical transport of dust $j_z = \rho v_z$ with a friction time of $\tauS$ is given by the sum of sedimentation under gravitational acceleration $g_z$ and diffusion of strength $D$:
\begin{equation}
j_z = \tauS g_z \rho - D \partial_z \rho.
\end{equation}
For the equilibrium solution $j_z = 0$ 
this leads to the differential equation:
\begin{equation}
\frac{\tau}{D} g_z  = \partial_z \ln \rho.
\end{equation}
Taking another derivative in $z$ and assuming ${\tau}$ and ${D}$ to be independent of height, leads to
\begin{equation}
\frac{\tau}{D} \nabla g_z  = - \frac{\tau}{D} \nabla^2 \Phi = \partial_z^2 ln \rho.
\end{equation}
For a plane parallel self gravitating dust layer we find $\nabla^2 \Phi = - 4 \pi G \rho$
and by expressing the dust density in units of the Hill density we get 
\begin{equation}
\frac{\rho}{\rho_c} =  - \frac{D}{\tau}\frac{1}{4 \pi G \rho_{\rm Hill}} \frac{\rho_{\rm Hill}}{\rho_c}\partial^2_z \ln{\rho},
\label{eq:plane}
\end{equation}
for a given central dust density of $\rho_c$.
The solution is not a Gaussian, but a hyperbolic function \begin{equation}
\rho(z) =\rho_c \cosh^{-2}{\frac{z}{\sqrt{2} \hlcrit}}.
\label{eq:cosh}
\end{equation}
Here, the characteristic scale height $\hlcrit$ expands upon the critical scale length $\lcrit$ that follows from the time scale argument \citep[see e.g.,][]{KlahrSchreiber2020a}, i.e.,
\begin{equation}
\hlcrit = \frac{1}{3}  \sqrt{\frac{\rho_{\rm Hill}}{\rho_c}} \sqrt{\frac{\delta}{St}} H = \sqrt{\frac{\rho_{\rm Hill}}{\rho_c}} l_c,
\label{eq:firsthlc}
\end{equation}
The expression $D/\tau$ has the units of a velocity $a$ squared:
\begin{equation}
a^2 = \frac{D}{\tau} = \frac{\delta}{St} \cs^2.
\label{eq:a2}
\end{equation}
Thus $h_p = a / \Omega$ determines the pressure scale height in the same way as does the speed of sound for the scale height of the gas $H = \cs / \Omega$. We therefor refer to $a$ as the equivalent speed of sound for pebbles, which especially for small pebbles is very different from the global r.m.s.\ velocity of pebbles. For $\stokes \rightarrow 0$ the r.m.s.\ velocity of pebbles will approach the turbulent velocity of the gas $\approx \sqrt{\delta} \cs$, whereas $a$ will rise up to the speed of sound (see Equation \ref{eq:asound} for the full expression allowing for $\stokes < \delta$).

\section{Justification of the Toomre Ansatz}
\label{sec:B}
\corrected{
In Section \ref{sec:3} we use a linearised version of simplified hydro dynamic equations (Equations \ref{eq:BTrho} - \ref{eq:BTvphi}) to derive a Toomre stability criterion for dust under the influence of turbulent diffusion. Here we derive and justify our chosen ansatz.}

The full set of equations the coupled evolution of gas ($\mathbf{u},\rhoGas$) and dust ($\mathbf{v},\rho$) under self gravity $\nabla^2 \Phi_s = - 4 \pi G \left(\rhoGas + \rho\right)$ and stellar gravity $\Phi_\odot$, thus $\Phi = \Phi_\odot + \Phi_s$ as we also solve it in our Pencil Code simulations is given by
\begin{eqnarray}
\partial_t \rhoGas + \nabla \cdot (\rhoGas \mathbf{u}) &=& 0, \label{eq:gascont}\\
\partial_t \rho + \nabla \cdot (\rho \mathbf{v}) &=& 0,\\
\partial_t \mathbf{u} + (\mathbf{u} \cdot \nabla)\mathbf{u} &=& -\frac{1}{\rhoGas} \nabla P -\frac{\rho}{\tauS \rhoGas} \left(\mathbf{u} - \mathbf{v} \right) - \nabla \Phi,\label{eq:gaseul} \\
\partial_t \mathbf{v} + (\mathbf{v} \cdot \nabla)\mathbf{v} &=&  -\frac{1}{\tau} \left(\mathbf{v} - \mathbf{u} \right) - \nabla \Phi.
\end{eqnarray}
For the system studied in the paper, we know that using the right parameters SI will generate turbulence in the gas, but that the gas stays quasi in-compressible and all mean velocities of the gas average out to zero $\mathbf{u} = \mathbf{u'}$ once we average in time over one turbulent correlation time $\tau_\mathrm{t}$, which is even longer than the coupling time $\tau_\mathrm{t}$ for $\stokes < 1$ and we can directly ignore Equations \ref{eq:gascont} and \ref{eq:gaseul} and set 
$\mathbf{u'}^2 = u^2$ and $\rhoGas = \mathrm{const}$.
The diffusion of the dust is driven by the turbulent gas motions and depends on the r.m.s.\ velocity of the gas but also on the correlation time of the turbulence, the turbulent spectrum and the Stokes number as well as the dust to gas ratio of the particles \citep{YoudinLithwick2007}. This means, that just based on the r.m.s.\ velocities of the particles $v^2$, one cannot calculate the diffusivity. One first has to determine the diffusivity $D_{\mathrm{g}}$ of the gas itself, which as described by \citep{YoudinLithwick2007} has to be determined form the power spectrum of the turbulence $E_{\mathrm{g}}(\omega)$ with $\omega$ the frequencies of the Fourier modes as 
\begin{equation}
D_{\mathrm{g}} = \int_0^\infty d \tauS \int_{-\infty}^{\infty} d\omega E_{\mathrm{g}}(\omega) e^{-i \omega \tau} = <u^2> \tau_\mathrm{t}.
\end{equation}
From that the diffusivity for particles of a given stokes number can be determined, yet all under the assumption that turbulence and diffusion is isotropic. 
\citet{JohansenYoudin2007} have shown that in their simulations, which are the basis for our simulations in the present paper, that the directly measured radial and vertical diffusivity of small particles and an estimate based on the particle r.m.s.\ velocities in radial and vertical direction multiplied by the determined correlation time is in agreement "within a factor of a few random velocities":
\begin{equation}
   D_{x,z} = v^2_{x,z} \tau_\mathrm{t}.
   \label{eq:DXZ}
\end{equation}
Thus we skipped the analysis of turbulence in terms of correlation times and velocity spectrum for this paper and used exclusively the diffusivities that we measure, either by tracking particles for unstratified conditions or by measuring the dust scale height in stratified cases.

If we now use the measured pebble diffusivities $D$ to determine the diffusion flux of pebbles
$\rho v_{\mathrm{diff}} = - D \rhoGas \nabla \frac{\rho}{\rhoGas}$ following Fick's law, as addition to the dust equations we get the set of equations used in \citet{Youdin2011} for the derivation of the secular gravitational instability:
\begin{eqnarray}
\partial_t \rho + \nabla \cdot (\rho (\mathbf{v} - \frac{D}{\rho} \nabla \rho)) &=& 0, \label{eq:B1.f1}\\
\partial_t \mathbf{{v}} + (\mathbf{{v}} \cdot \nabla)\mathbf{{v}} &=& -\frac{1}{\tau} \left(\mathbf{v} - \mathbf{u} \right) - \nabla \Phi,\label{eq:B1.f2}
\end{eqnarray}
where we already drop $\rhoGas$, as it is spatially constant in our simulations. 
The momentum Equation has not changed here by the diffusion equation.
This may lead to errors for the coupled system as noted by \citet{Tominaga2019}, for instance angular momentum is not conserved. We will avoid this by redefining our velocity as combined transport velocity of drift and diffusion part $\mathbf{v}^*$.
\correctedn{For a stationary case with constant density gradients we can define:}
$\mathbf{v}^* := \mathbf{v} - \frac{D}{\rho} \nabla \rho$, 
\correctedn{but for a dynamic situation we have to solve the following set of equations to determine the evolution of $\mathbf{v}^*$.}
\begin{eqnarray}
\partial_t \rho + \nabla \cdot (\rho \mathbf{v}^*) &=& 0,\\
\partial_t \mathbf{v}^* + (\mathbf{v}^* \cdot \nabla)\mathbf{v}^* &=& X -\frac{1}{\tau} \left(\mathbf{v}^* - \mathbf{u} \right) - \nabla \Phi.
\end{eqnarray}
Here $X$ is the source term that drives the diffusion flux, which we can determine from the equilibrium situation $\mathbf{v}^*=\mathbf{u}=\partial_t \mathbf{u} = \partial_t \mathbf{v}^* = 0$ of diffusion vs.\ sedimentation. This is the same ansatz that Einstein used for his derivation of Brownian Motion \citep{Einstein1905}.
Without Diffusion particles will settle with velocity $v = - \tauS \nabla \Phi$ and this velocity has to be balanced by diffusion $\tauS \rho \nabla \Phi  = - D \nabla \rho$. We can now replace $\nabla \Phi$ and get:
\begin{eqnarray}
X &=& - \frac{D}{\tauS \rho} \nabla \rho.
\end{eqnarray}
Thus our final dynamical Equation is:
\begin{eqnarray}
\partial_t \rho + \nabla \cdot (\rho \mathbf{v}^*) &=& 0, \label{eq:final1}\\
\partial_t \mathbf{v}^* + (\mathbf{v}^* \cdot \nabla)\mathbf{v}^* &=& - \frac{D}{\tau} \frac{\nabla \rho}{\rho} -\frac{1}{\tau} \left(\mathbf{v}^* - \mathbf{u} \right) - \nabla \Phi.\label{eq:final2}
\end{eqnarray}
Now the coupling of dust and gas considers the complete mass flux (advective and diffusive) and thus also the fictitious forces are correct and angular momentum conservation is automatically achieved, \correctedn{without adding explicit diffusion flux to the momentum equations as was proposed by \citet{Tominaga2019}. Note the mathematical difference between adding the "instantaneous" diffusive flux to the momentum equation to adding a diffusion pressure to the momentum equation, that will drive and regulate the diffusive flux under conservation of momenta.}

We can use $\mathbf{v}^*$ then also for the momentum equation of the gas:
\begin{eqnarray}
\partial_t \mathbf{u} + (\mathbf{u} \cdot \nabla)\mathbf{u} &=& -\frac{1}{\rhoGas} \nabla P + \frac{D}{\tau} \frac{\nabla \rho}{\rhoGas} -\frac{\rho}{\tauS \rhoGas} \left(\mathbf{u} - \mathbf{v}^* \right) - \nabla \Phi,\label{eq:gaseul2}
\end{eqnarray}
where we also added the momentum that flows into pebble diffusion.
Note, that when stratification of the gas has to be taken into account, then in the above derivations the gradient of density for the pebbles $\nabla \rho$ has to be replaced by $\rhoGas \nabla \rho/\rhoGas$.

In our new set of equations (\ref{eq:final1} and \ref{eq:final2}) we can for instance study the onset of diffusion. Assume that $\mathbf{u} = 0$ and that $\mathbf{v}^*(t=0) = 0$. Then we find 
\begin{eqnarray}
\partial_t \mathbf{v}^* &=& - \frac{D \frac{\nabla \rho}{\rho} - \mathbf{v}^*}{\tau},
\end{eqnarray}
that within one stopping time the diffusive flux will reach its equilibrium value.

In accordance to this derivation our linearised Equations for the Toomre stability ansatz would be \correctedn{similar yet not identical to} \citep{Ward1976,Ward2000,Youdin2011}, \correctedn{as the diffusion term is in the momentum and not in the continuity equation}:
\begin{eqnarray}
\frac{1}{\Sigma}\partial_t \Sigma' + \partial_r  v_r' &=& 0,\\
\partial_t v_r' - 2 \Omega v_\phi' &=& -\frac{1}{\Sigma} \frac{D}{\tau} \partial_r \Sigma' - \partial_r \Phi' - \frac{v_r'}{\tau},\\
\partial_t v_\phi' + \frac{\kappa^2}{2 \Omega} v_r' &=& - \frac{v_\phi'}{\tau_\phi}.
\end{eqnarray}
Using as always $\Phi' = - 2 \pi \Sigma' G / |k|$ in the dispersion relations for plane waves $e^{i\left(\omega t - k r\right)}$ gives \begin{eqnarray}
\left(\omega - \frac{i}{\tau_\phi}\right)\left[\Omega^2 - 2 \pi \Sigma G |k| + \frac{\delta}{\stokes} c_s^2 k^2 - \omega \left(\omega - \frac{i}{\tau}\right)\right] + \frac{i}{\tau_\phi} \Omega^2 =0 .
\label{eq:SGI_DISP}
\end{eqnarray}
\correctedn{This leads to the secular gravitational instability including diffusion as extensively discussed in \citet{Youdin2011}. That work found overstable (oscillatory growing) modes because diffusion in the continuity equation leads to additional complex terms in the dispersion relation. If one additionally adds the diffusion to the left hand side of the momentum equation as done in \citet{Tominaga2019}, the additional complex terms are avoided and the instability is not oscillating \citep{Tominaga2020}.
Yet our new ansatz is much simpler and directly shows that $\omega$ is not oscillatory, i.e. has no real parts. If we determine the growthrates $\gamma = i \omega$ we get:}
\begin{eqnarray}
\left(\gamma + \frac{1}{\tau_\phi}\right)\left[\Omega^2 - 2 \pi \Sigma G |k| + \frac{\delta}{\stokes} c_s^2 k^2 + \gamma \left(\gamma + \frac{1}{\tau}\right)\right] - \frac{1}{\tau_\phi} \Omega^2 =0 .
\label{eq:SGI_DISP2}
\end{eqnarray}
\correctedn{If we assume $\tau_\phi = \tau \ll 1$ in Equation \ref{eq:SGI_DISP2} we receive the growthrates for the secular gravitational instability as:}
\begin{eqnarray}
\gamma = \tau 
\left(2 \pi \Sigma G |k| - \frac{\delta}{\stokes} c_s^2 k^2 \right).
\label{eq:SGI_GROWTH}
\end{eqnarray}
\correctedn{Thus the treatment of diffusion as a driving term in the momentum equation has several benefits. For A it is much simpler in terms of mathematics. For B it leads directly to monotonical growing modes and C the growth rates are written in the same manner as if $D/\tau$ was the rms velocity of the pebbles as in \citet{ChiangYoudin2010}.}

\begin{figure}
    \centering
    \includegraphics[width = 0.5\linewidth]{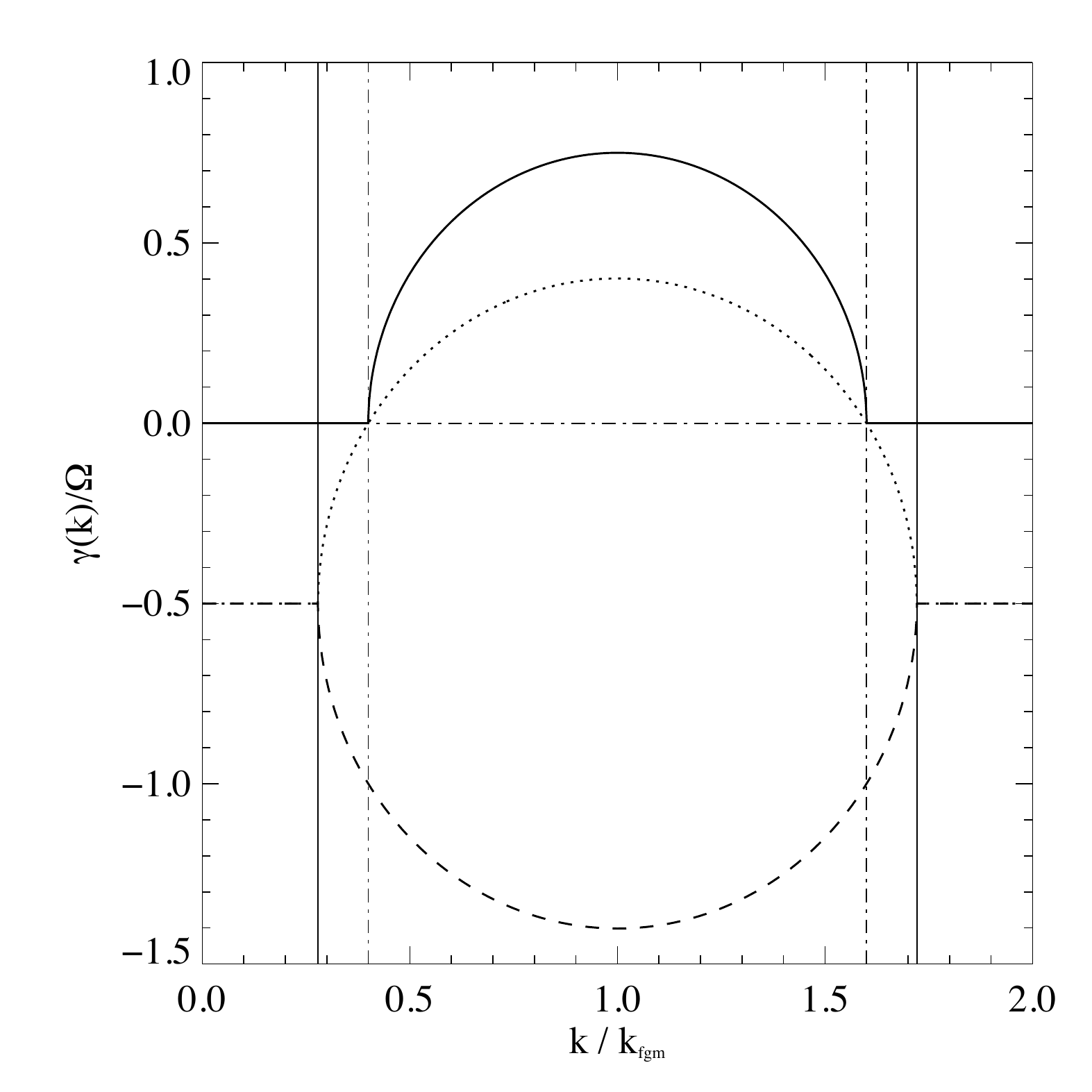}
    \caption{Growthrates $\gamma(k)$ in units of orbital frequency $\Omega$ as function of fastest growing mode $k_\mathrm{fgm}$ for $Q = 0.8$. The solid line is the normal Toomre growth rate ($\stokes = \infty$) and the dotted line is for $\stokes = 1$. In both cases the fastest, the smallest and the largest growing wave numbers ($\gamma > 0$) are identical with $k_\mathrm{min} = 0.4 k_\mathrm{fgm}$ and $k_\mathrm{max} = 1.6 k_\mathrm{fgm}$ (vertical dash-dotted line). We also plot the second solution for $\gamma$, which is always negative (dashed line). Note the  $k_{-} = 0.28 k_\mathrm{fgm}$ and $k_{+} = 1.72 k_\mathrm{fgm}$ for strongest damping and transition from exponential decay to damped oscillations (vertical solid line).}
    \label{fig:5}
\end{figure}

\begin{figure}
    \centering
    \includegraphics[width = 0.5\linewidth]{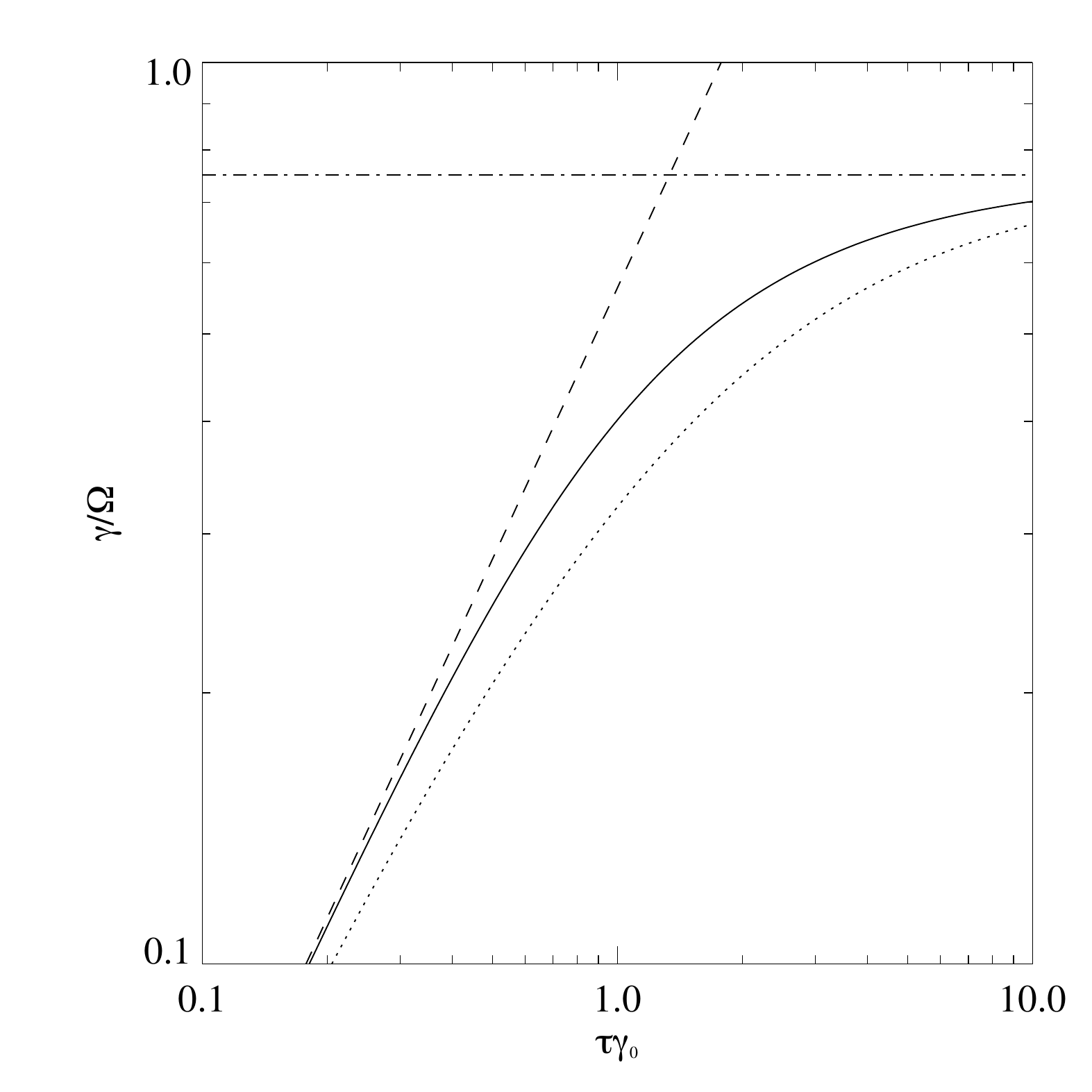}
    \caption{Growthrates for the fastest growing mode $\gamma$ in units of orbital frequency $\Omega$ as function of friction time $\tauS$ for $Q = 0.8$. The solid line is the damped Toomre growth rate and the dashed line the approximation for $\stokes \gamma < 1$. We also show the undamped growth rate $\gamma_0$ as horizontal dash-dotted line and an analytic fit for the growth-rates as dotted line.}
    \label{fig:6}
\end{figure}

For the situation studied in this paper, when the dust starts to exceed the local gas mass, it is no longer justified that the gas can help the dust to loose its excess in angular momentum as provided by term $- \frac{v_\phi'}{\tau_\phi}$. We keep the second friction related term $- \frac{v_r'}{\tau}$ as the gas is still able to slow down radial contraction by its pressure. Then the new dispersion relation is simpler:
\begin{eqnarray}
\omega \left(\omega - \frac{i}{\tauS}\right) = \frac{\delta}{\stokes} c_s^2 k^2 + \Omega^2 - 2 \pi \Sigma G |k| = \omega(k)_0^2,
 \label{eq:Toomre_full}
\end{eqnarray}
where we define the right hand side as the classical $\omega_0^2$ without gas friction.
The minimum of $\omega(k)_0^2$  can be found via $\mathrm{d}\omega(k)_0^2/\mathrm{d}k = 0$ and determines in the classical Toomre result the fastest growing mode:
\begin{eqnarray}
k_{\rm fgm} = \frac{\pi \Sigma G}{\frac{\delta}{\stokes} c_s^2},
\end{eqnarray}
and as the reader can easily prove, this is still the fastest growing mode if friction is included.
If we solve Equation \ref{eq:Toomre_full} for $\gamma = i \omega$ we find:
\begin{eqnarray}
\gamma(k) \left(\gamma(k) + \frac{1}{\tauS}\right) = - \omega(k)_0^2,
 \label{eq:Toomre_nice}
\end{eqnarray}
Thus $\omega_0 = 0$ implies $\gamma_1 = 0$ plus an exponential decaying solution $\gamma_2 = -\frac{1}{\tau}$.
The extrema for $\gamma$ can be found by taking the derivative with respect to $k$ on both sides of Equation\ \ref{eq:Toomre_nice}:
\begin{eqnarray}
 \left(2 \gamma + \frac{1}{\tauS} \right) \frac{\mathrm{d} \gamma}{\mathrm{d} k} = - \frac{\mathrm{d} \omega_0^2}{\mathrm{d} k} = 0,
\end{eqnarray}
and for whatever $k$ produces a minimum for $\omega^2_0$ (largest growthrate) gives $\gamma$ a maximum.
Interestingly, there are also additional extrema for:
\begin{eqnarray}
\gamma(k) = - \frac{1}{2 \tau},
\end{eqnarray}
which define the wave numbers $k_{\pm}$ for strongest damping and transition from exponential decay to damped oscillations.
In Figure \ref{fig:5} we plot growth rates for linear self gravity modes for $Q=0.8$ and a Stokes Number of 
$\stokes = 1$. The growth rates are always smaller than for the undamped case ($\stokes = \infty$). Yet, the maximum and the roots are the same for $\gamma(k)$ as for $\gamma_0(k)$.
Thus the Toomre value for the onset of instability is independent of $\stokes$:
\begin{eqnarray}
Q = \frac{\sqrt{\frac{\delta}{\stokes}} c_s \Omega}{\pi G \Sigma},
\end{eqnarray}
only the normalized growth rates $\gamma = - i \omega $ are diminished by $\stokes$. 
\begin{eqnarray}
\gamma = \sqrt{ \gamma_0^2 + \frac{1}{4 \tau^2}} - \frac{1}{2 \tau} \approx \gamma_0 \frac{\tauS \gamma_0}{1 + \tauS \gamma_0}.
\end{eqnarray}
In effect, the growth rates are diminished proportional to $\stokes$ with respect to the growth rates of infinite Stokes number $\gamma_0$ if $\gamma_0 \stokes < 1$ (see Figure \ref{fig:6}). 

As this slow down of growth for the linear mode is beyond what we have in mind for the main part of the paper, and we also showed that the Toomre stability criterion itself including the range of unstable wave numbers is independent on $\stokes$, plus we are stable in our simulations anyway, we omit the friction term $- \frac{v_r'}{\tau}$ in the main part of the paper (see Equations \ref{eq:BTvr} and \ref{eq:BTvphi}) for simplicity.
Nevertheless, for a more detailed study of Toomre instability of a pebble layer in a turbulent disk, these equations may come handy.

\section{Sound Waves for pebbles?}
\label{sec:C}
\correctedn{In the above sections we already argued that the "pseudo speed of sound" $a$ of pebbles describes their resistance against compression, just like a thermal pressure would do $P = \rho a^2$.} 

\correctedn{But in how far is $a$ really a propagation speed for waves and what is the benefit in defining diffusion as an effective pressure instead of adding it to the continuity equation, besides the above mentioned angular momentum conservation?}

\correctedn{Thus it is helpful to study the above dispersion relation (Equation \ref{eq:Toomre_full}) for the force free one-dimensional case:}
\begin{eqnarray}
\omega \left(\omega - \frac{i \Omega}{\stokes}\right) = \frac{\delta}{\stokes} c_s^2 k^2.
 \label{eq:Sound_full}
\end{eqnarray}
In case of well coupled pebbles $\stokes \rightarrow 0$ we find the classical diffusion solution:
\begin{eqnarray}
\omega  = i D k^2,
 \label{eq:onlyDiff}
\end{eqnarray}
but the general case has the solutions:
\begin{eqnarray}
\omega_{1,2} = \frac{i}{2 \tau} \pm \sqrt{\frac{\delta}{\stokes} c_s^2 k^2 - \frac{1}{4 \tau^2}}.
 \label{eq:Sound_full2}
\end{eqnarray}
Thus, as long as 
\begin{eqnarray}
k \leq k_c = \frac{1}{2 \sqrt{\delta \stokes} H}
\label{eq:critkc}
\end{eqnarray}
there are no oscillatory motions, just viscous diffusion. 
Only if $k > k_c$ then we find indeed damped sound waves as solution. And those sound waves approach indeed the propagation speed of $a = \sqrt{\frac{\delta}{\stokes}} c_s$ for $k \gg k_c$.
While this is a mathematical valid solution, these waves are probably not essential for pebbles accumulations, as they will decay in half a coupling time of the pebbles. Also in reality it would be impossible to maintain a constant diffusivity $\delta$ when going to smaller and smaller wavelength. In other words, for very large $k$ or steep gradients the diffusion flux will be limited by the r.m.s.\ speed of the pebbles.

Nevertheless, our defined "pseudo sound speed" $a$ is actually the propagation speed, of very quickly decaying pebble density fluctuations of pebbles. 

What would have happened if we had done this analysis starting with diffusion only in the continuity equation? Using Equations \ref{eq:B1.f1} and \ref{eq:B1.f2} leads to a dispersion relation,
\begin{eqnarray}
\left(\omega^* - i k^2 D \right) \left(\omega^* - \frac{i}{\tauS}\right) = 0,
\label{eq:Sound_fullalt}
\end{eqnarray}
which in comparison to Equation \ref{eq:Sound_full} contains an additional term:
\begin{eqnarray}
\omega^* \left(\omega^* - \frac{i}{\tauS} - i k^2 D\right) = \frac{\delta}{\stokes} c_s^2 k^2,
 \label{eq:Sound_fullalt2}
\end{eqnarray}
with the solutions:
\begin{eqnarray}
\omega^*_{1,2} = \frac{i}{2} \left(\frac{1}{\tau} + k^2 D\right) \pm \sqrt{\frac{\delta}{\stokes} c_s^2 k^2 - \frac{1}{4} \left(\frac{1}{\tau} + k^2 D \right)^2}.
 \label{eq:Sound_full2s}
\end{eqnarray}
\correctedn{The result is similar, yet does not allow for an oscillatory solution as for $a k \tau = 1$ the root completely vanishes and for any other $k$ the argument in under the root is negative. 
So in that case there is no wave-number that allows for a pebble sound wave.}

\correctedn{So what solution is physical? Both solutions are identical for vanishing diffusivity $D = 0$. But for very large pebbles with $1 / \tau = 0$, the second solution $\omega^*$ still prescribes decaying perturbations, even so those objects would decouple from the gas and not be subject to diffusion any more and thus there is no reason why those perturbations should be damped. Also from a physical side it is clear why there are no sound waves and this the non-conservation of pebble momentum, via the instantaneous diffusion flux in the continuity equations.}

\correctedn{Nevertheless, are sound waves for wavenumbers as defined in Equation \ref{eq:critkc} a realistic scenario for pebbles in turbulent diffusion? And the surprising mathematical answer is yes. As pointed out by \citet{Tominaga2019} and others, a diffusive pebble flux also contains momentum. Thus when pebbles diffuse with respect to a density gradient they carry net momentum. And this allows (for very large wave numbers) for an overshooting, i.e.\ the momentum can still be larger than zero when the density gradient vanishes and creating a new local density extremum. Not for long, as the wave decays on a coupling time, yet formally this constitutes a sound wave.}

\correctedn{The pebble sound speed $a$ is smaller than the speed of sound of the gas $c$ yet can be larger than the actual pebble r.m.s.\ velocity. But this is a general problem with Ficks law and is not specific for treating diffusion in the momentum equation rather than in the continuity equation. As soon as density gradients become too steep (same as very large $k$ values), the diffusion can produce larger fluxes than what would be possible by the actual r.m.s.\ velocity of turbulence, respectively pebbles. Therefor a flux limiter is some times necessary to be implemented for diffusion over strong gradients.}

\correctedn{We can discuss a special case here, which even may be able to be tested numerically. For stopping times on the order the correlation time of the turbulence, or roughly speaking a stokes number of unity we know that the r.m.s.\ speed $v_\mathrm{r.m.s.}^2 = \delta c_s^2$ equals the pebble sound speed $a^2$. Then the critical wave length is}
\begin{eqnarray}
\lambda_c = \frac{2 \pi}{k_c} = 4 \pi \sqrt{\delta} H,
\end{eqnarray}
\correctedn{which is slightly larger than the mixing length of turbulence of $L = \sqrt{\delta} H$. Thus it could be possible that a sinusoidal perturbation of $\stokes = 1$ pebbles would indeed show an oscillatory wavelike behaviour, which could be tested at least in numerical experiments.}

\correctedn{In conclusion, mathematical speaking, diffusion allows for sound waves at large wave numbers $k$ and large stokes numbers.
Yet for small pebbles $\stokes < 1$ and large wave numbers, respectively short distances, the diffusion ansatz is likely to break down. So for the pebble sizes studied in the present paper, there are no pebble sound waves under turbulent diffusion. But still $a$ is the sound speed of the pebbles under turbulent diffusion in our formalism for critically damped (non-oscillatory) solutions.}





\end{document}